\documentclass{aa}
\usepackage{graphicx}
\usepackage{natbib,twoopt}
\usepackage[breaklinks=true]{hyperref} 
\bibpunct{(}{)}{;}{a}{}{,} 
\makeatletter
\newcommandtwoopt{\citeads}[3][][]{\href{http://adsabs.harvard.edu/abs/#3}%
{\def\hyper@linkstart##1##2{}%
\let\hyper@linkend\@empty\citealp[#1][#2]{#3}}}
\newcommandtwoopt{\citepads}[3][][]{\href{http://adsabs.harvard.edu/abs/#3}%
{\def\hyper@linkstart##1##2{}%
\let\hyper@linkend\@empty\citep[#1][#2]{#3}}}
\newcommandtwoopt{\citetads}[3][][]{\href{http://adsabs.harvard.edu/abs/#3}%
{\def\hyper@linkstart##1##2{}%
\let\hyper@linkend\@empty\citet[#1][#2]{#3}}}
\newcommandtwoopt{\citeyearads}[3][][]%
{\href{http://adsabs.harvard.edu/abs/#3}
{\def\hyper@linkstart##1##2{}%
\let\hyper@linkend\@empty\citeyear[#1][#2]{#3}}}
\makeatother
\usepackage[varg]{txfonts}
\usepackage{hyperref}
\begin{document}

\title{Empirical instability strip for classical Cepheids II.  The Small Magellanic Cloud galaxy}

   \subtitle{}

   \author{Felipe Espinoza-Arancibia\inst{1}
          \and
          Bogumił Pilecki\inst{1}
          \and
          Matylda Łukaszewicz\inst{2}
          }

   \institute{$^1$ Nicolaus Copernicus Astronomical Center, Polish Academy of Sciences, Bartycka 18, 00-716 Warsaw, Poland\\
   $^2$ Astronomical Observatory, University of Warsaw, Al. Ujazdowskie 4, 00-478 Warsaw, Poland\\
              \email{fespinoza@camk.edu.pl}
             }

   \date{Received xxxx; accepted xxxx}

 
  \abstract
  %
   {}
   {This study aims to determine empirical intrinsic edges of the classical Cepheids instability strip (IS) in the Small Magellanic Cloud (SMC) galaxy, considering various effects that alter its shape, and compare them with theoretical models and other galaxies.}
   {We used the data of classical fundamental-mode (F) and first-overtone mode (1O) SMC Cepheids from the OGLE-IV variable star catalog, with the final cleaned sample including 2388 F and 1560 1O Cepheids. The IS borders are determined by tracing the edges of the color distribution along the strip. Based on that, and using evolutionary tracks, the IS crossing times are then calculated.}
   {We obtained the blue and red edges of the IS in V- and I-photometric bands and in the HR diagram, and detected breaks at periods between 1.4 and 3 days. Interestingly, the central SMC Cepheids are redder than those located farther away. A comparison with existing theoretical models showed good agreement for the blue edge and significant differences for the red edge. We also found that the IS of the SMC is wider than that of the Large Magellanic Cloud (LMC), with its red edge being redder despite its lower metallicity. The analysis of crossing times showed that the expected number of Cepheids as a function of period agrees with the observed distribution for $P > 1$ days but differs for $ P < 1$ days.}
   {Slope changes along the SMC IS borders are most likely explained by the distribution of metallicity. The behavior of the blue loops at the SMC metallicity is not consistent with observations, and at the LMC metallicity, the blue loops are too short for lower-mass stars. A comparison of theoretical edges with our empirical ISs imposes constraints on the models and enables the identification of valid ones. Based on the positions of the breaks, our study also suggests that fundamental-mode Cepheids with periods longer than 3 days should be used for distance determination.}


   \maketitle

\section{Introduction}

The classical instability strip (IS) is a region in the Hertzsprung-Russell diagram (HRD) characterized by the presence of partially ionized layers of H and He in intermediate-mass stars. These layers play a crucial role in the excitation of radial pulsations through the $\kappa$ and $\gamma$ mechanisms, leading to the formation of classical Cepheids \citep[see, e.g.,][]{catelan2015}. These stars exhibit a relationship between their pulsation period and luminosity, which places them as crucial objects for determining extragalactic distances \cite[for a recent review, see][]{Bono2024}. Additionally, classical Cepheids (hereafter Cepheids) are valuable for verifying stellar evolution and pulsation theories \cite[some recent studies are, e.g.,][]{Hocde2024, Marconi2024, Stuck2025, Deka2025}. Typically, Cepheids first cross the IS during the H-shell burning phase after leaving the main-sequence phase. The second and third crossings occur during the He-core burning phase, commonly known as the blue loop. This evolutionary phase is highly sensitive to metallicity and adopted input physics, such as convective overshooting, nuclear reactions, and rotation \citep[see, e.g.,][]{2004Xua, Walmswell2015, Espinoza2022, Zhao2023, Ziolkowska2024}.

Numerous theoretical investigations have examined the effects of various physical properties on the IS. Among them, \citet[][ and references therein]{Marconi2005} used nonlinear convective pulsation models with different metal and helium abundances. The authors noted that the IS edges shift to redder colors as metallicity increases (at fixed helium abundance) and that the red edge increases its effective temperature as helium abundance increases (at fixed metallicity). \cite{Anderson2016} investigated the impact of rotation on Cepheid models with varying metal content. Their results indicated that the blue edge of the IS is not significantly affected by rotation, whereas the red edge shows a slight shift toward red as rotation increases. Recently, \citeauthor{Somma2024} (2024, and references therein) analyzed the topology of the IS using updated opacity tables on nonlinear pulsation calculations, obtaining good agreement with their previous results and supporting the trend of the IS to get redder as metallicity increases. \cite{Deka2024} investigated the effects of the free parameters of the convective model of the pulsation functionality of the code Modules for Experiments in Stellar Astrophysics \citep[MESA;][]{Paxton2011, Paxton2013, Paxton2015, Paxton2018, Paxton2019, Jermyn2023}, Radial Stellar Pulsations \citep[RSP;][]{Smolec2008,Paxton2019}. Their results indicated that enabling additional physical processes in the convective model (by using more complex RSP parameter sets) shifts the edges of the IS toward redder colors. Recently, \cite{Khan2025} presented a comprehensive study of metallicity effects on Cepheid Period-luminosity relations using synthetic Cepheid populations computed with the Geneva stellar evolution models and the SYCLIST tool. In particular, they obtained the IS borders for metallicities representative of the Sun, the Large Magellanic Cloud (LMC), and the Small Magellanic Cloud (SMC), which were in good agreement with previous studies.

On the other hand, there are significantly fewer empirical studies that are focused on the IS properties. Among the initial efforts, \cite{PelLub1978, Fernie1990, Turner2001} obtained an empirical IS of our Galaxy. \cite{Tammann2003} obtained the IS of the Galaxy and the Magellanic Clouds based on period-luminosity (P-L) and period-color relations. They found differences in the slopes of the ISs between the three galaxies. \cite{Sandage2004, Sandage2009} studied samples of Cepheids from the LMC and the SMC. They identified breaks in the P-L relations at specific pulsation periods: 10 days for the LMC and 2.5 days for the SMC. The break observed in the LMC's P-L relation was also observed at the IS edges. More recently, in Paper I \citep{Espinoza2024}, we computed an empirical IS of the LMC Cepheids, using a sample of 2058 fundamental-mode (F) and 1387 first-overtone mode (1O) Cepheids. We reported a break in the edges of the IS, located at the pulsation period of about 3 days. We compared our empirical boundaries with theoretical ones from the literature and found good agreement.

Following the same method as in Paper I, we aim to extend our results to the SMC by obtaining an empirical intrinsic IS for the Cepheids in that galaxy, using the most recent Cepheid catalogs available. The comparison between empirical and theoretical ISs can be used to constrain different physical processes that affect the position of the IS edges. Moreover, we can empirically study the effect of metallicity on the IS by comparing the results of this work with those of Paper I for the LMC.

The outline of this paper is as follows: Sect.~\ref{sec:sample} describes the sample selection procedure; Sect.~\ref{sec:ISborders} describes the method used to obtain the IS borders; Sect.~\ref{sec:discussion} presents a discussion of our results, including a comparison with theoretical evolutionary tracks and ISs published in the literature. Finally, Sect.~\ref{sec:conclusion} presents our conclusions.

\section{Sample selection}\label{sec:sample}
\begin{figure}
\centering
\includegraphics[width=\hsize]{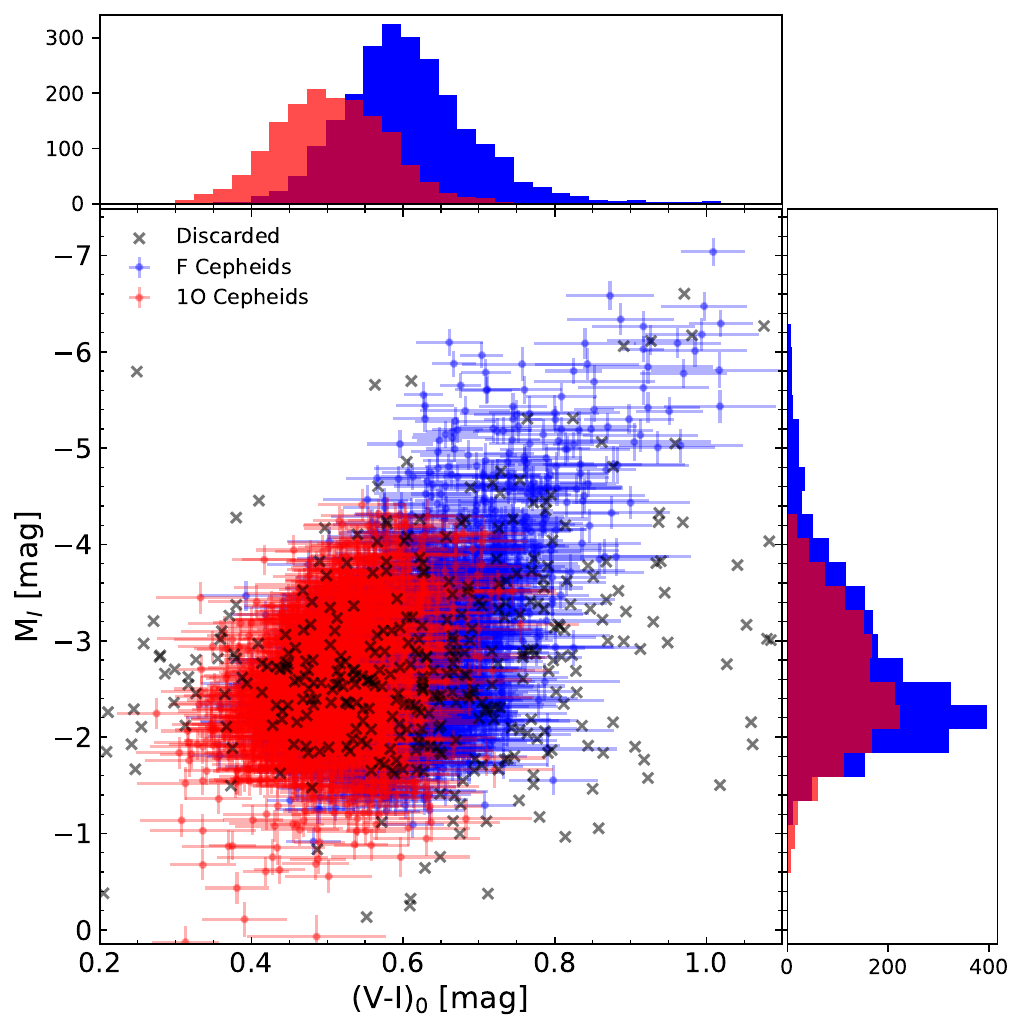}
\caption{CMD of the final sample of F (blue) and 1O (red) SMC Cepheids. The objects discarded in our cleaning procedure are shown as crosses. Distributions of the intrinsic color (V-I)$_0$ and absolute magnitude $M_I$ are shown in the upper and right subpanels, respectively.}
\label{fig:CMD}
\end{figure}

We use data of F and 1O Cepheids in the SMC from the OGLE-IV variable stars catalog \citep{2017AcA....67..103S}, and followed the same cleaning procedure as in Paper I. In short, we removed outliers from the reddening-free period-luminosity relation (also known as period-Wesenheit or PW relation), to get rid of binary Cepheids with luminous companions \citep{Pilecki2021} or with otherwise affected luminosity (blended, etc.). In addition, we discarded Cepheids that presented remarks in the OGLE-IV catalog, and objects that deviated more than three sigma from the relation between the magnitude residuals of the I-band P–L relation and the corresponding residual of the P–W relation \citep{Madore2017}. We computed the reddening correction for each Cepheid using the reddening map by \citet{Skowron2021} and the coefficients of relative extinction by \citet{Schlegel1998} to calculate the intrinsic $(V-I)_0$ color of the sample\footnote{We considered $A_I/A_V=0.594$, and $E(V-I)=1.238~E(B-V)$.}. As a final step in the cleaning procedure, we discarded objects with reddening uncertainties above the $95^{\rm th}$ percentile of the reddening error distribution. We estimated individual distances to each Cepheid using the refined geometrical model of the SMC computed by \citet{Breuval2024} and used them to compute absolute I-band magnitudes. We discuss the geometrical shape of the SMC in Section~\ref{subsec:geometry}. The distribution in the CMD of the final sample of 2388 F and 1560 1O Cepheids is shown in Fig.~\ref{fig:CMD}.

\section{Instability strip borders}\label{sec:ISborders}

To achieve our goal of determining the intrinsic empirical boundaries of the IS, we must account for any factors that could influence its width. Only after making these corrections can we ensure that our results are comparable to theoretical models. In Section~\ref{sec:sample}, we described the criteria used to exclude outliers from our sample. The most significant remaining factor is the impact of reddening uncertainties. These uncertainties increase color scatter among Cepheids, thereby widening the IS. Note that, while photometric uncertainty also affects the IS width, its impact is approximately one order of magnitude smaller.

We performed the same method introduced in Paper I (Section 3 therein) to obtain intrinsic IS edges. First, we binned the whole Cepheids sample by I-band absolute magnitude. Most bins contain 200 stars, while the faintest contain around 100. We then determined each bin's initial blue and red IS edges by locating the $1^{\rm st}$ and $99^{\rm th}$ percentiles of the color distribution. We then shifted each Cepheid's intrinsic color by a random value drawn from a normal distribution with a standard deviation equal to the measured color uncertainty. Such a procedure results in some of them falling outside the initial edges. Subsequently, we separately counted the number of Cepheids that fell behind the red and blue edges. We repeated this process 10000 times and computed the median of the distribution of these numbers, namely $n_{\rm blue}$, $n_{\rm red}$. The final blue and red IS positions of each bin were obtained by moving the initial edges inward the IS by $n_{\rm blue}$ and $n_{\rm red}$ stars, respectively.

The Cepheids in the SMC have lower mean reddening than those in the LMC. In the LMC, the sample-average reddening was approximately $0.12$ mag, whereas in the SMC it is $0.06$ mag. Nevertheless, the reddening map from \cite{Skowron2021} shows similar average uncertainties of about $0.06$ mag for both galaxies. This means that for the LMC and SMC, the effect of the IS widening due to the inaccurate dereddening of the sample, which affects the IS topology as discussed in Section~\ref{subsec:break}, should be comparable.

\begin{figure*}
\sidecaption
\includegraphics[width=0.6\linewidth]{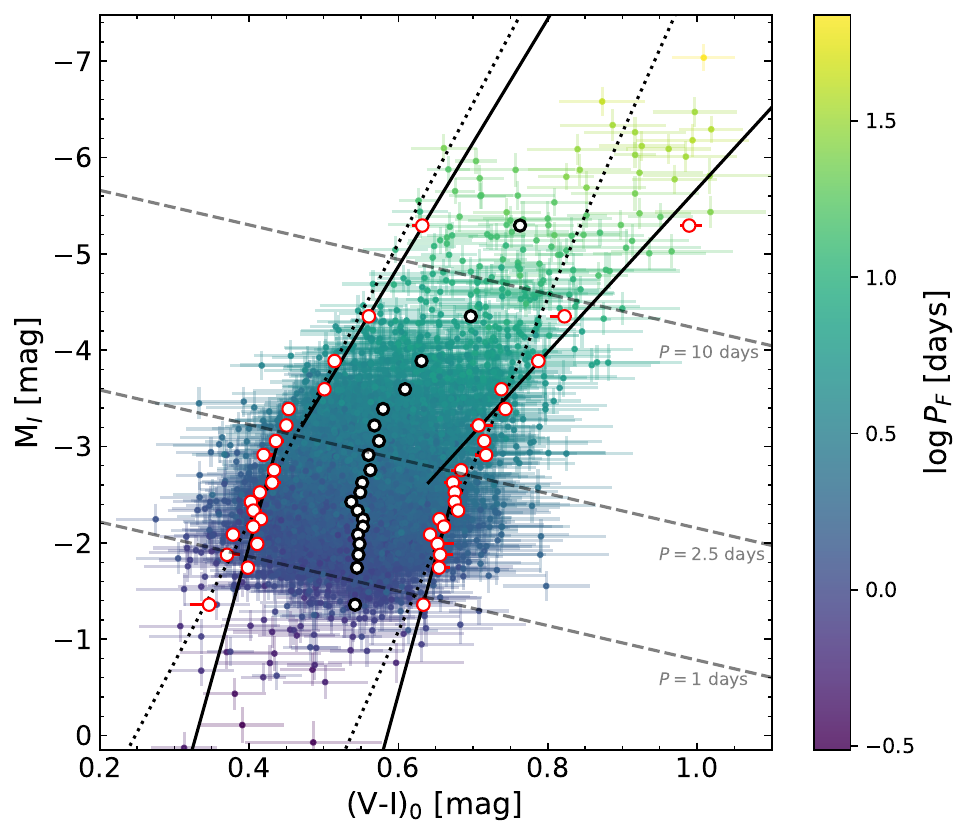}
\caption{CMD of F and 1O SMC Cepheids. The boundaries of our empirical IS are shown as red empty circles. The median intrinsic colors of each bin are shown as black empty circles. The upper and lower parts of the fitted red and blue edges are shown as black solid lines. The edges of the wedge-shaped IS are shown as dotted lines. Periods for these stars are shown with a color gradient. For 1O Cepheids, periods were fundamentalized. Dashed lines of constant periods are overplotted.}
\label{fig:ISfull}
\end{figure*}

The computed IS, including F and 1O Cepheids, are shown in Fig.~\ref{fig:ISfull} as red circles. Additionally, the median intrinsic colors $(V-I)_0$ of each sample bin are shown as black circles. In this figure and throughout this paper, the periods of the 1O Cepheids were fundamentalized using equation (1c)\footnote{$P_{\rm F}/P_{\rm 1O} = 1.356 + 0.068\log P_{\rm 1O}.$} of \cite{Pilecki2024} and we overplotted constant period lines for $1$, $3$, and $10$ days. These lines were calculated using the period-luminosity-color (PLC) relation computed using the I-band absolute magnitude and intrinsic color of the sample, and the periods provided in the OGLE catalog. Similar to our results in Paper I, we observe a change in slope of the calculated edges between faint and bright Cepheids close to the three-day constant period line. This change can also be noticed in the median intrinsic color along the IS (hereafter, median IS). To determine the break position in the IS edges, we use piecewise regression analysis \citep{Muggeo2003} implemented in Python by \citet{Pilgrim2021}. This technique employs a statistical hypothesis test to assess the significance of the break by testing whether at least one breakpoint exists against the null hypothesis of no breakpoints. We checked every fit of the IS boundaries, and in all cases the null hypothesis was rejected at a significance level below $1\%$ (i.e., the probability of observing the data if the null hypothesis were true is below 1\%). A larger number of breakpoints was not necessary; therefore, to describe the blue and red IS edges, we chose the model composed of two segments. Nevertheless, we kept the simplified wedge-shaped IS model (with no breaks) for comparison. Both IS models are shown in Fig.~\ref{fig:ISfull} and the coefficients of the fitted edges with 1-sigma uncertainties are listed in the Appendix (Table \ref{tab:table1}). We emphasize that the wedge-shaped IS is just a crude approximation that has significant systematic deviations from the determined IS shape. Therefore, its formally determined uncertainties are not statistically meaningful, and are not provided. A discussion about the possible origin of the identified breaks can be found in Section \ref{subsec:break}.

\begin{figure*}
\resizebox{\hsize}{!}
{\includegraphics{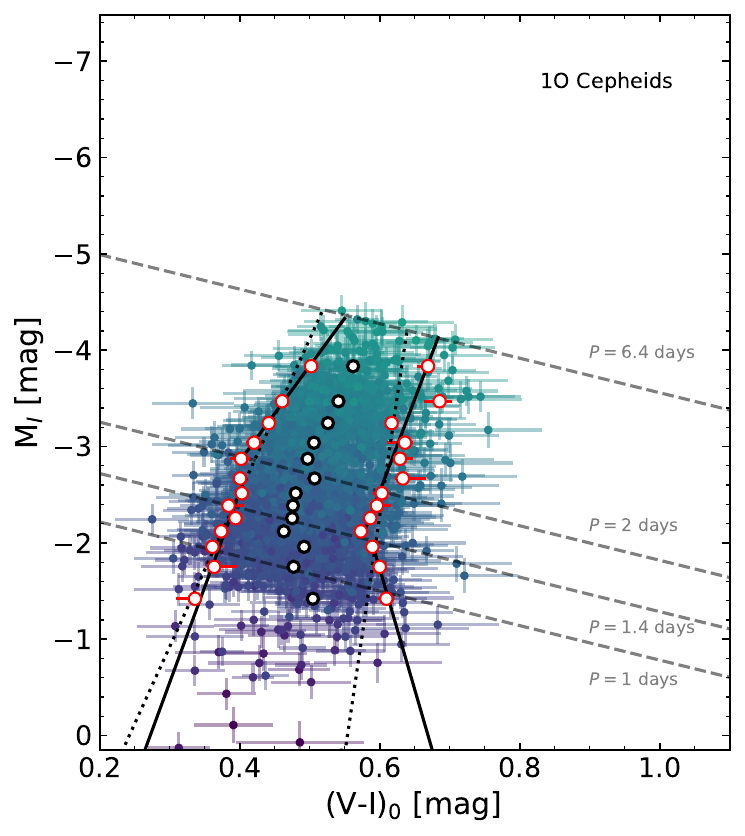}
\includegraphics{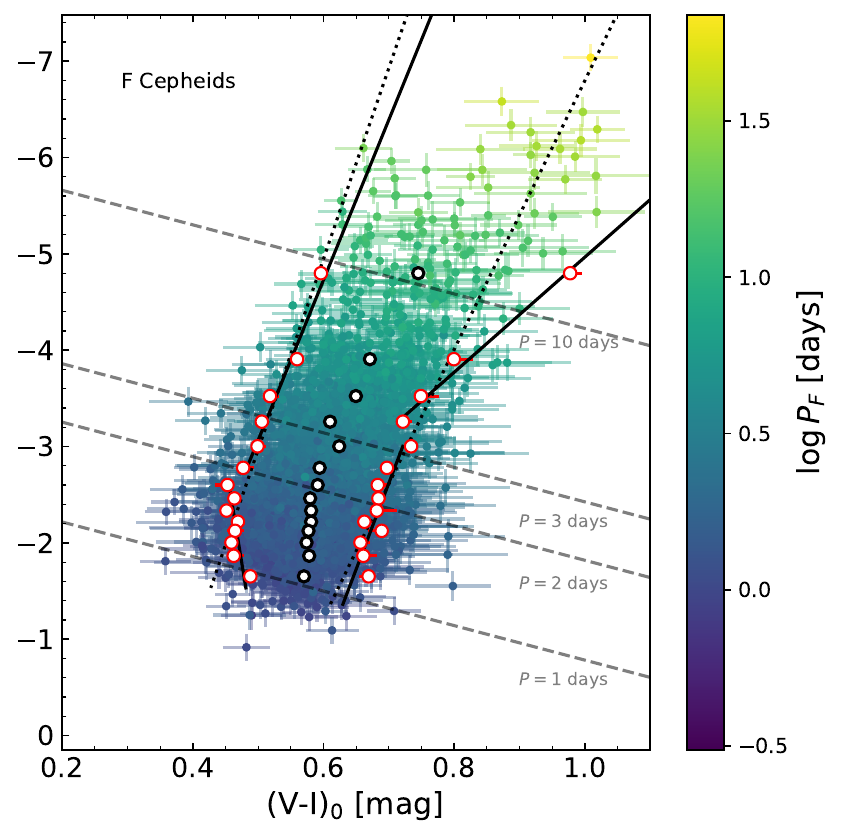}}
\caption{Same as Fig.~\ref{fig:ISfull} but independently for 1O- (left panel) and F (right panel) Cepheids.}
\label{fig:IS1OF}
\end{figure*}

Apart from the IS for the full sample, we also determined its edges separately for F and 1O Cepheids. The borders obtained for these subsamples are shown in Fig.~\ref{fig:IS1OF} as red empty circles\footnote{Data points describing the edges in a machine-readable form are available at \url{https://doi.org/10.5281/zenodo.17277341}}. 
Changes in the IS edge slope were detected in both sets, although at different pulsation periods compared with each other and with the full sample. As before, we fitted and presented in the figure both the two-segment linear models and a wedge-shaped IS. The corresponding coefficients for the IS edges are given in Table \ref{tab:table1}.

\section{Discussion} \label{sec:discussion}

\subsection{Topology of the IS boundaries in (V-I) color}\label{subsec:break}

Previous studies have reported breaks in the P-L relation of F and 1O Cepheids in the Magellanic Clouds. \cite{Bauer1999} noticed a slope change in the P-L relation for the SMC F Cepheids with periods shorter than two days, which was later confirmed by \cite{Sharpee2002, Sandage2009}. In a recent detailed analysis, \citet{Kurbah2025} identified multiple breaks in the period-amplitude, P-L, and amplitude-color relations of both F and FO Cepheids in the Magellanic Clouds. One of these breaks is located at a pulsation period of $2.5$ days, which is close to the findings of previous studies. As shown graphically by \citet{Madore1991}, the IS edges and the P-L relation are projections of the PLC relation onto the $\log L$ vs $\log T$, and the $\log L$ vs $\log P$ plane, respectively. Therefore, features observed in one of these projections may also be visible in the others.

We observe changes in the slopes of the IS for our full sample, and also in the individual samples of F and 1O Cepheids. For our full sample, the break occurs at fundamentalized periods of around $2.5$ days, specifically at absolute I-band magnitudes of $-3.08$ mag and $-2.59$ mag for the blue and red edges, respectively. This is similar to what we found for the LMC in Paper I, and consistent with the findings of \cite{Kurbah2025}. On the other hand, for the individual F and 1O samples, we found changes in the slopes of the blue and red edges of the IS at different pulsation periods. For 1O Cepheids, the blue edge shows a break at a pulsation period of around 2 days (fundamentalized). However, the red edge shows the same feature at a shorter period of around 1.4 days (fundamentalized). The locations of these breaks in I-band absolute magnitude are $-2.74$ mag and $-2.13$ mag for the blue and red edge, respectively. In the case of F Cepheids, their blue edge shows a change in slope at a pulsating period of around 2 days, meanwhile, the red edge shows the same feature at around 3 days, similar to the full sample, which is expected since the red edge of the entire sample is dominated by F Cepheids, as seen in Fig~\ref{fig:CMD}. These breaks are located at $-2.59$ and $-3.38$ I-band absolute magnitudes. Similar conclusions can be drawn for each sample when tracing the median along the IS. For the full sample, the break in the median IS is around 2.5 days, while for the F and 1O subsamples, the change in slope occurs between the breaks at the blue and red edges for a given pulsation mode. This indicates that the position of the break shifts gradually as we move from one edge to another.

In Paper I, we describe a break in the IS boundaries for LMC Cepheids with periods of around $3$ days. A possible explanation for this feature is the depopulation of second and third-crossing Cepheids in the faint part of the IS. Since the extension of the blue loops decreases as the mass and period of Cepheids decrease, low-mass Cepheids would spend less time inside the IS, or they would not enter at all. This implies that Cepheids fainter than the break are likely on their first crossing of the IS. To test if the depopulation scenario could also produce features in the IS of the SMC, we used the stellar evolution code MESA (version r22.11.1) to calculate evolutionary tracks for non-rotating stars covering the mass range from 2 to 7~ $\mathrm{M}_{\sun}$ in steps of $0.1~\mathrm{M}_{\sun}$. The lower end of this mass range was extended, compared to the one used in Paper I, to consider all low-mass tracks that still crossed entirely or partially the SMC IS. We adopted  $Z=0.001$, $0.002$, and $0.003$ as representative metallicities for SMC stars \citep{Choudhury2018}. These tracks assume the solar mixture from \cite{Grevesse1998}. We use a solar-calibrated mixing length parameter of $\alpha_{\rm mlt} = 1.9$. For the convective boundaries, we use the predictive mixing scheme described in \cite{Paxton2018}. \cite{Smolec2023} studied the behavior of blue loops as a function of core and envelope overshooting. They showed that the larger the envelope overshooting, the longer the blue loop. Based on this and after exploring parameters, we considered the exponential core and envelope overshooting with parameters $f = 0.015$ and $f = 0.024$, respectively. For this set of parameters, the extension of the blue loops qualitatively matches the observed IS, in particular, the resulting blue loops completely cross the IS. To account for mass loss during the evolution on the red giant branch (RGB), we use the \cite{Reimers1975} prescription with a scaling factor $\eta_R = 0.1$. 

\begin{figure*}
\centering
\includegraphics[width=.45\textwidth]{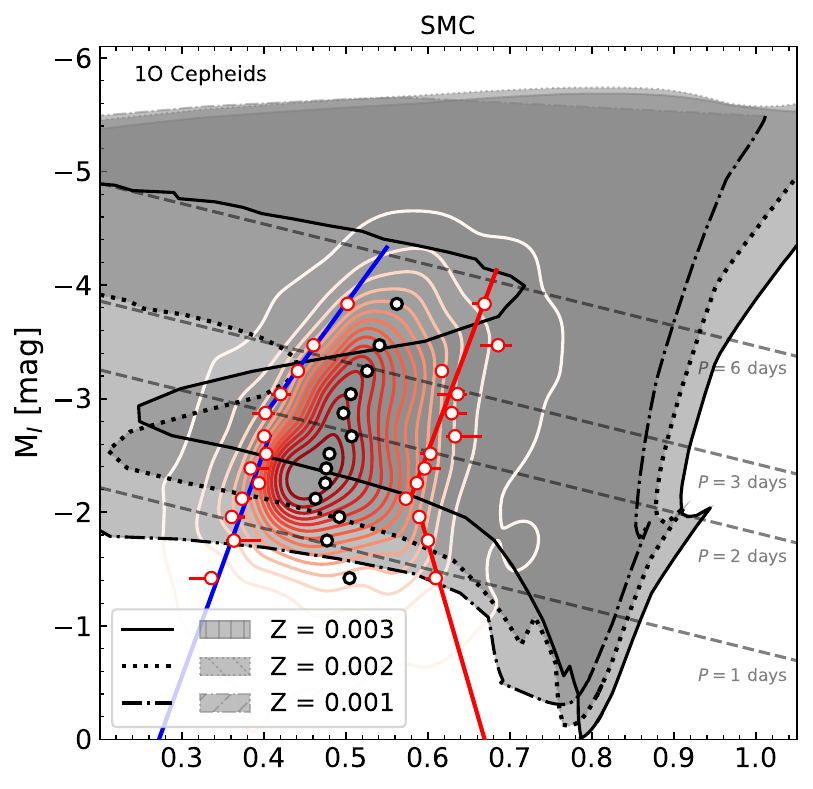}
\includegraphics[width=.43\textwidth]{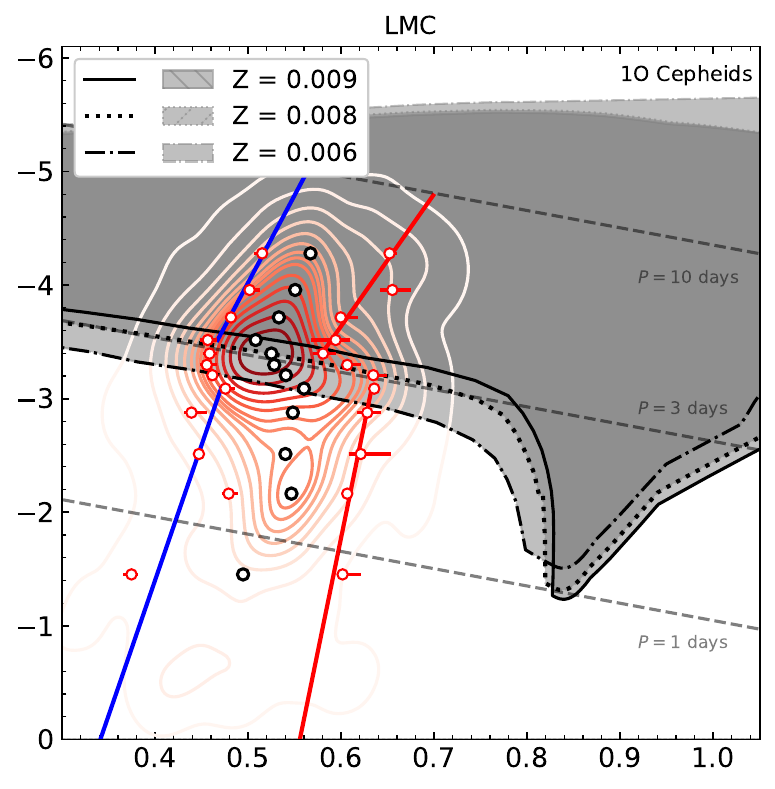}
\includegraphics[width=.45\textwidth]{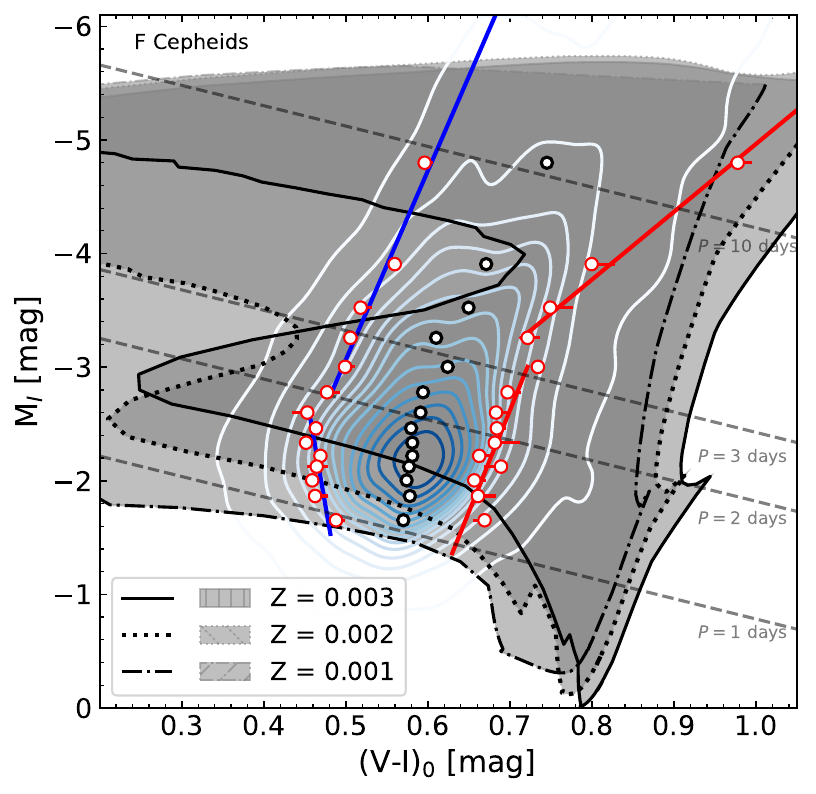}
\includegraphics[width=.43\textwidth]{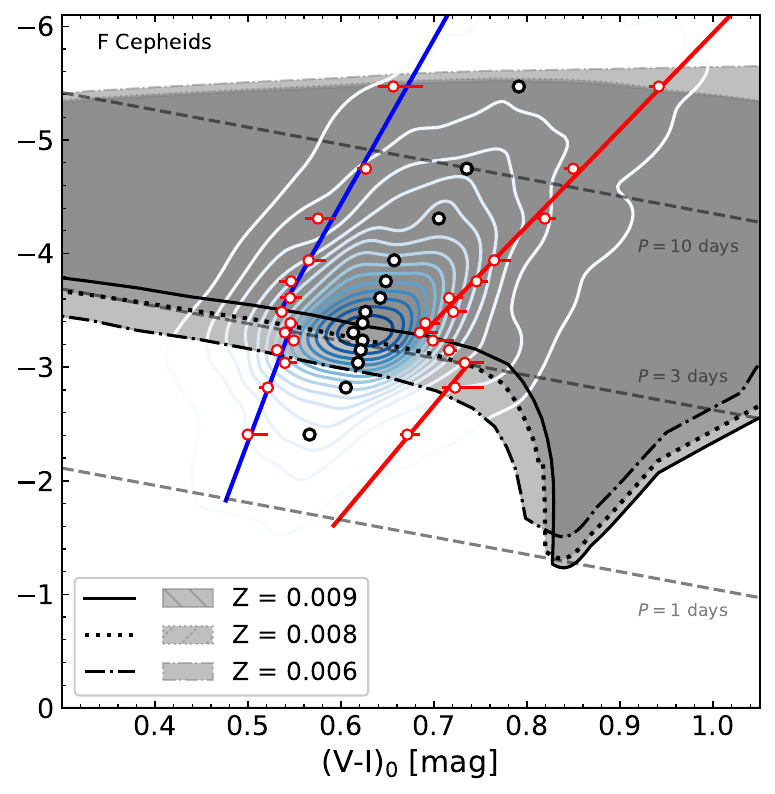}
\caption{CMD showing empirical IS edges (red empty circles) separately for SMC (left panels) and LMC (right panels), 1O (upper panels) and F (lower panels) Cepheids. Black empty circles represent the median value of the intrinsic color of each bin of the corresponding sample. Fits for the blue and red edges, considering changes in slope at different pulsation periods, are shown as solid blue and red lines, respectively. Gray-shaded areas mark the blue loop extent (delimited by its bluest extreme and the tip of the RGB) for evolutionary tracks with representative $Z$ for the SMC and LMC galaxies. The gray areas are created using tracks from $2$ to $7$ M$_\sun$ for the SMC, and $3$ to $7$ M$_\sun$ for the LMC. The upper limits for the areas are defined by the evolutionary tracks for $7$ M$_{\sun}$ and would extend further up if tracks for higher masses are considered.}
\label{fig:IS-F1O-MCs}
\end{figure*}

The comparisons between evolutionary tracks and our empirical IS borders for the SMC (left panels) and LMC (Paper I, right panels) are shown in Fig.~\ref{fig:IS-F1O-MCs}. In this figure, we show the extent of the blue loops (shaded areas) for each adopted metallicity. The tip of the RGB delimits these areas on the red side and the bluest extreme of the blue loop on the blue side. The upper limits are artificial and are defined by the evolutionary tracks for 7~$\mathrm{M}_{\sun}$-stars with different metallicities. Additionally, we show contours of the Cepheids' density distributions of the 1O and F samples for both galaxies. The constant period lines were obtained independently for each galaxy. A discussion of the comparison between the SMC and LMC results can be found in Section~\ref{subsec:compLMC}. 

In the case of the SMC, the model blue loops are particularly sensitive to metallicity, and their extensions vary in a more complex manner than in the LMC. For $Z=0.002$ and $0.003$, the extension ``oscillates'' around the IS as the mass of the star increases. For these models, the loops enter the IS at a Cepheid mass of $2.7$ and $3.1$ M$_\sun$, respectively. These mass limits for blue loops are much lower than in the LMC (Paper I). Subsequently, their blue loop extension reaches a local maximum, gradually decreasing to a minimum extension at masses of $3.8$ and $4.5$ M$_\sun$, respectively. We tried lower values of envelope overshoot, but the ``oscillatory'' effect was even more pronounced and the blue loops were clearly inconsistent with the data. On the other hand, the evolutionary tracks with $Z=0.001$ show a more regular behavior since the extension of their blue loops increases as a function of stellar mass. These models enter the IS at a Cepheid mass of $2.3$ M$_\sun$, and their blue loops cover our F and 1O samples almost entirely.

The interpretation of results for the SMC is notably more intricate than that for the LMC due to a more complex influence of metallicity. Contrary to the LMC, the SMC model blue loops (except for $Z=0.003$) cover almost completely the range of observed periods. This means that the depopulation scenario, which alters the slope of the IS, cannot be tested appropriately because the first and subsequent crossing Cepheids are mixed. However, the lower blue loop boundary for the highest metallicity correlates with the break for 1O Cepheids and the blue edge for F Cepheids. As the lower boundary for the lowest metallicity approaches the shortest periods in the sample, what we see as a change in slope may be an extended transition phase to a depopulated area. Such a transition phase is also present in the LMC, but because the lower blue loop boundaries for different Z values have a much smaller brightness spread, it is much tighter.

Another complication in interpreting the SMC data is the ``oscillation'' of the blue loop extensions mentioned above, which is absent in the LMC evolutionary tracks. This phenomenon suggests a significant decrease in the number of 2nd- and 3rd-crossing Cepheids with pulsation periods of around $3$ to $10$ days and a metallicity of $Z=0.003$. This effect is not observed in our empirical results, which indicates there may be an issue with the choice of particular input physics that could affect blue loops in the evolutionary models. This effect has also been reported in the literature. As shown in \cite{Smolec2023}, the extent of the blue loops shows nonlinear behavior as a function of mass for specific metallicity values and does not cross the IS for some masses (generally tracks with $4$ or $5$ M$_\sun$). Furthermore, evolutionary tracks computed with the PARSEC v2.0 code \citep{Costa2025}, adopting a metallicity of $Z=0.002$, and BaSTI \citep{Hidalgo2018} tracks adopting $Z=0.003$ show the same blue loop behavior. This could indicate that evolutionary models in general require fine-tuning for the loops at pulsation periods between $3$ and $10$ days to cross the IS. On the other hand, the distribution of metallicities of our Cepheids sample could be closer to lower metallicities than $Z=0.003$. A more accurate understanding of the SMC metallicity distribution could let us better understand the IS's topology.

\subsection{Theoretical crossing times of the ISs}\label{subsec:xing}

\begin{figure*}
\centering
\includegraphics[width=.9\hsize]{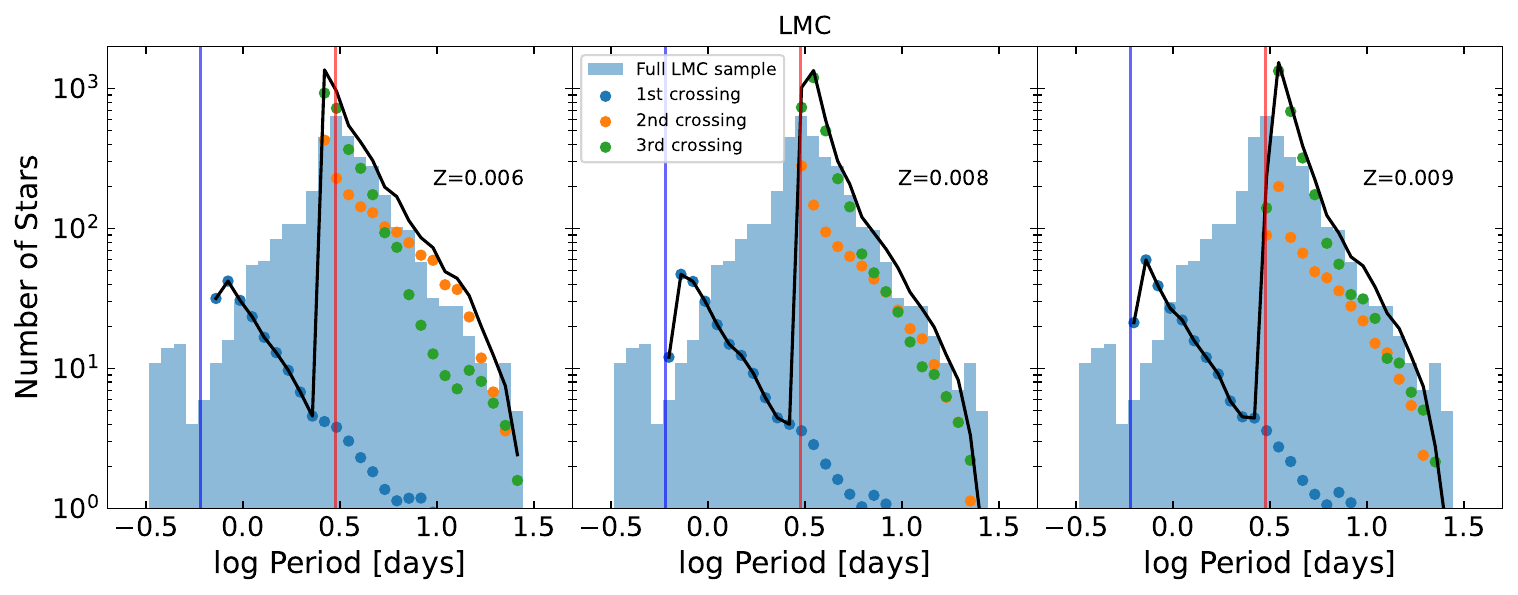}
\includegraphics[width=.9\hsize]{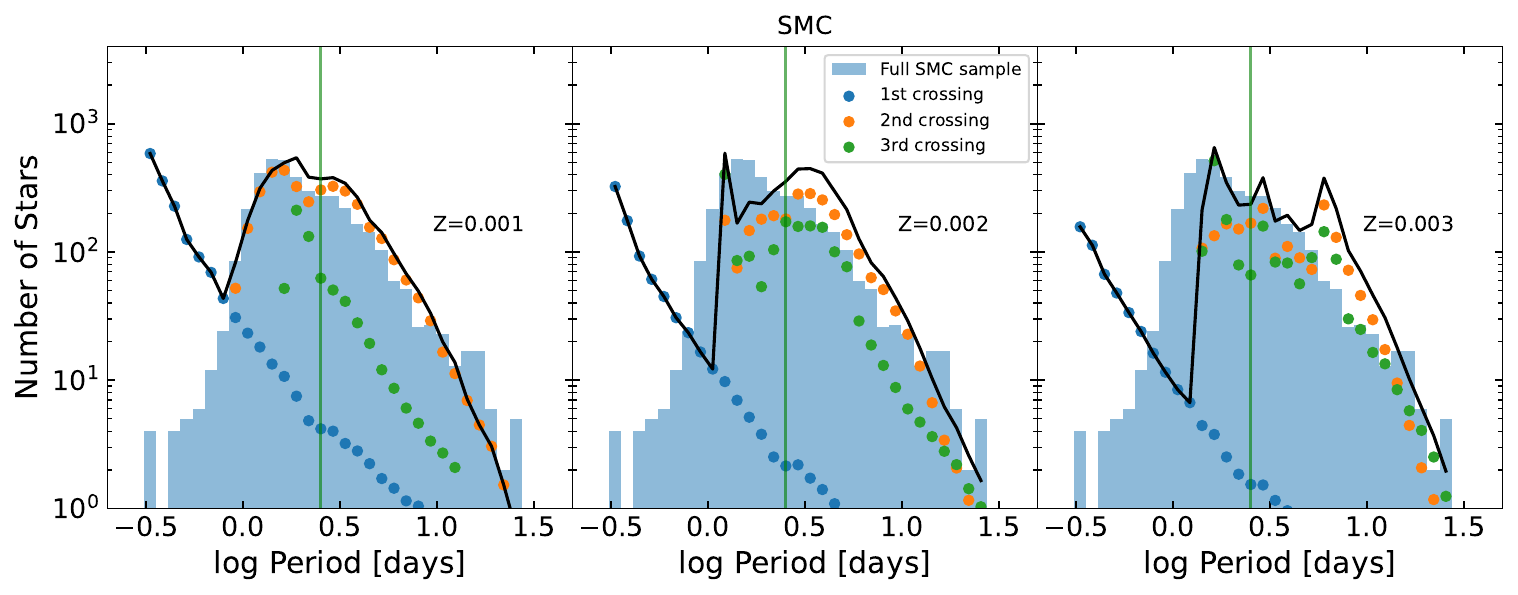}
\caption{Histogram of the number of LMC (upper panels), and SMC (lower panels) Cepheids, including F and 1O, as a function of fundamentalized pulsating period. Colored points show the expected number of Cepheids on each bin for different IS crossings, calculated using Eq.~\ref{eq:1}. The sum of the expected numbers of stars for all IS crossings is shown as a black solid line. Vertical blue, red and green lines mark pulsation periods of $0.6$ (the shortest period of our LMC models), $3$, and $2.5$ days, respectively.}
\label{fig:xing}
\end{figure*}

Using the evolutionary and pulsation models computed employing MESA and RSP, we estimated the expected number of Cepheids as a function of pulsation period, IS crossing, and metallicity, defined as follows:
\begin{equation}\label{eq:1}
    \mathcal{N}_{F,1O} (P,Xing,Z) = N_{F,1O} \frac{\tau_{F,1O
    }(P,Xing,Z)}{ \sum_{P,Xing}\tau_{F,1O}(P,Xing,Z)}\xi(m)\Delta m,
\end{equation}
where $N_{F,1O}$ is the number of F and 1O Cepheids in our sample, and $\tau_{F,1O}(P,Xing,Z)$ is the time that a Cepheid model with a pulsation period $P$ and a metallicity $Z$ spent inside a crossing ($Xing$) of our full-sample IS. The factor $\xi(m)\Delta m$ is \citet{Salpeter1955} initial mass function (IMF), where the dependency on mass was changed to dependency on period, using a period-mass relation computed from RSP models. This estimate assumes a constant star formation rate for the SMC. We compared the expected number of Cepheids with the observed number in our sample in Fig.~\ref{fig:xing} for the LMC (upper panels) and SMC (lower panels) galaxies. For this analysis, 1O periods were fundamentalized, as mentioned in Section~\ref{sec:ISborders}. 

In the case of the LMC, the maximum of the Cepheid distribution is located at a pulsating period of 3 days. Then the number of stars gradually decreases as the period increases. For pulsating periods higher than 3 days, the expected number of Cepheids obtained for the second and third crossings is in good agreement with the observed number of stars. As mentioned in Paper I, for periods shorter than $3$ days, the blue loops of our tracks do not cross the IS. Therefore, the expected number of short-period Cepheids drops significantly to only the first-crossing models. Since we computed evolutionary tracks starting from $2$ M$_{\sun}$, our results are limited to periods longer than 0.6 days, i.e., to the range in which the blue solid line in the upper panels of Fig.~\ref{fig:xing} is plotted. 

A significant discrepancy between theoretical predictions and observations appears for Cepheids with pulsation periods shorter than 3 days. In the 1-3-day period range, the observed number of Cepheids decreases with period and, thus, with stellar mass. Conversely, theoretical models, primarily based on the first crossing of the IS, predict an increasing number of stars as the period shortens. This theoretical trend is influenced by two factors: a slower evolution rate for lower-mass stars, which increases their transit time across the IS, and the higher contribution of low-mass stars in the IMF.

Agreement between the expected and observed Cepheids is only achieved at a period of about 1 day. The observed excess of Cepheids between 1 and 3 days could potentially be explained by stars on a gradually narrowing blue loop that partially enter into the IS but do not cross it. However, current stellar evolution models fail to produce blue loops of sufficient extension to account for such objects in the LMC. An alternative scenario is that these Cepheids have lower metallicity than assumed in our models, as blue loops for more metal-poor stars may cross the IS at shorter periods.

For periods shorter than 1 day, the predicted number of Cepheids exceeds the observed number. For the shortest periods, however, the statistics of our sample are low, and the edges of the IS are poorly constrained (determined from only one sparsely populated bin). If all stars passing through the IS at a given brightness would pulsate, we would expect an increase in their number with decreasing mass (and shortening Cepheid period). Although some Cepheids exist in this zone, their number is much lower than predicted. This may indicate that, as mass decreases, the probability of exciting F or 1O pulsation modes decreases rapidly. On the other hand, this short-period region brings additional challenges for pulsation theory, which we use to determine the period of Cepheids passing the IS. Pulsation models presented by \citet[][and references therein]{Buchler2009} showed that these short-period Cepheids present multi-mode pulsations (F/1O or involving higher overtones), together with hysteresis (i.e., the pulsational state depends on the evolutionary path). In such cases, determining the final pulsation mode and period requires nonlinear calculations that might alter the expected period distribution of Cepheids shown in Fig.~\ref{fig:xing}. Nevertheless, such an analysis is beyond the scope of this paper.

In the case of the SMC galaxy, the peak of the Cepheid distribution occurs at significantly shorter periods than in the LMC, at around 1.4 days. There is close agreement between the observed and expected numbers of Cepheids with periods longer than $1$ day, particularly for models with $Z=0.001$. At higher metallicities, the expected number of stars differs from the observed number, likely due to the complex behavior of theoretical blue loops at such metallicities. Notably, the models for $Z=0.001$ also successfully reproduce the observed Cepheid counts in the 1-3-day period range, whereas for the LMC, substantial disagreement was observed in that range across all considered values of Z.

Although for the SMC, better agreement is obtained for Z=0.001, this does not mean its average metallicity is that low, as differences for this property between observations and theory for Cepheids are quite common \citep[see, e.g.,][]{Taormina2018, Deka2025}. Our theoretical value is still quite close to the one indicated by empirical studies, which is closer to $Z=0.002$ \citep{Breuval2024}. Actually, our data could help calibrate theoretical models in terms of metallicity.

For periods shorter than $1$ day, the expected number of stars differs significantly from the observed one.  Given the low number of observed Cepheids in this period range, the same factors that produce such discrepancies for the LMC can apply to the SMC.

\subsection{The geometry of the SMC}\label{subsec:geometry}

As shown by \citet{Jacyszyn2016}, Cepheids trace an elongated shape along the line of sight of the SMC galaxy, covering a distance range between 50 and 80 kpc. This can increase the uncertainties in distance determinations for these objects, potentially affecting their positions in the CMD and, in turn, the IS shape. To mitigate this effect, we adopted the refined planar model by \citet{Breuval2024} to correct the distance to each Cepheid. Despite applying this procedure, we explored whether the spatial distribution of the SMC Cepheids could still affect the shape of the IS borders.

\begin{figure}
\centering
\includegraphics[width=\hsize]{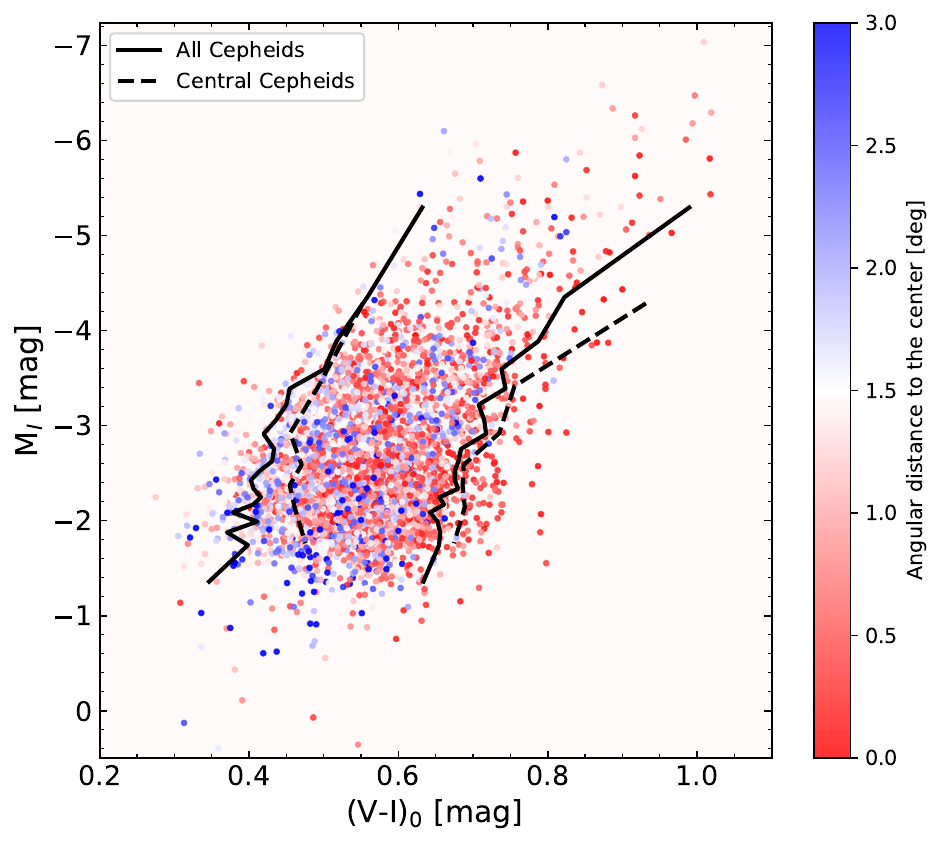}
\caption{CMD of F and 1O SMC Cepheids. Angular separation of Cepheids from the SMC center $(12^\circ.54,-73^\circ.11)$ is shown with a color gradient. The IS edges for the full sample and for only those within a $0.6$ degree radius from the SMC center are shown as solid and dashed lines, respectively.}
\label{fig:F1O_central}
\end{figure}

In Fig.~\ref{fig:F1O_central}, we show our full sample of Cepheids colored by their angular distance to the SMC center at $(12^\circ.54,-73^\circ.11)$ from \cite{Ripepi2017}. Superimposed, we compare the IS edges of the full sample with those determined for a subsample of central Cepheids located within $0.6$ deg from the SMC center. For clarity, here we join the points of the determined IS edges. Interestingly, these central Cepheids tend to be redder than those located farther away. The red edge is slightly, but systematically, shifted toward redder colors. For the blue edge, this depends on the luminosity: for fainter Cepheids, the difference is significant, but for brighter ones it decreases to zero. 

An inspection of the samples showed the mean intrinsic color uncertainties (which are dominated by the reddening uncertainties) are higher for central Cepheids compared to the complete sample, by 0.02 mag. Although this difference is significant, it should not ultimately affect the positions of the determined intrinsic edges. As mentioned in Section~\ref{sec:ISborders}, higher uncertainties may indeed lead to a wider IS, but our correction procedure should return it to its original state. Moreover, any residual effect could leave the IS either slightly narrower or wider, but not shifted, and not by the observed amount. Although, in principle, the edges can also be affected by the uncertainty of the distance correction for Cepheids located farther from the center, we also do not expect it to result in a systematic shift in color. 

The observed shift to redder colors may indicate that central Cepheids are more metal-rich than the average metallicity. Recently, \citet{Murray2024} showed that the SMC is composed of two substructures with different chemical compositions, separated by $5$ kpc along the line of sight. Nevertheless, the difference in metallicity within their sample of stars is relatively small, and it is unlikely that it would be responsible for the shifts in the IS edges that we have observed. Another possibility for shifting the whole IS to redder colors is an additional reddening for the central Cepheids, which is not accounted for in the reddening maps we used. This is quite likely, as the procedure used by \citet{Skowron2021} is calibrated in the galaxy's outer regions.

Actually, with the above test, we intended to check whether the IS for central Cepheids would be slightly narrower due to lower geometrical corrections to the absolute magnitudes. This, however, is not unambiguously observed.

\subsection{Comparison with the LMC ISs}\label{subsec:compLMC}

\begin{figure*}
\centering
\includegraphics[width=.49\hsize]{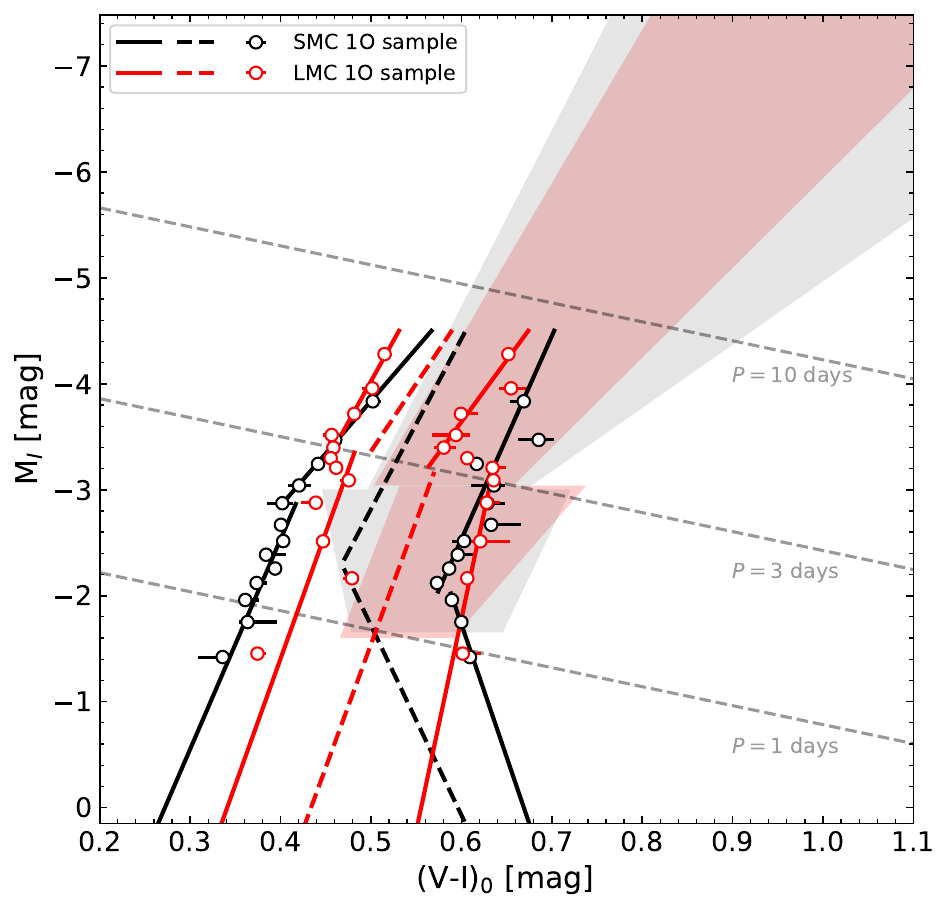}
\includegraphics[width=.47\hsize]{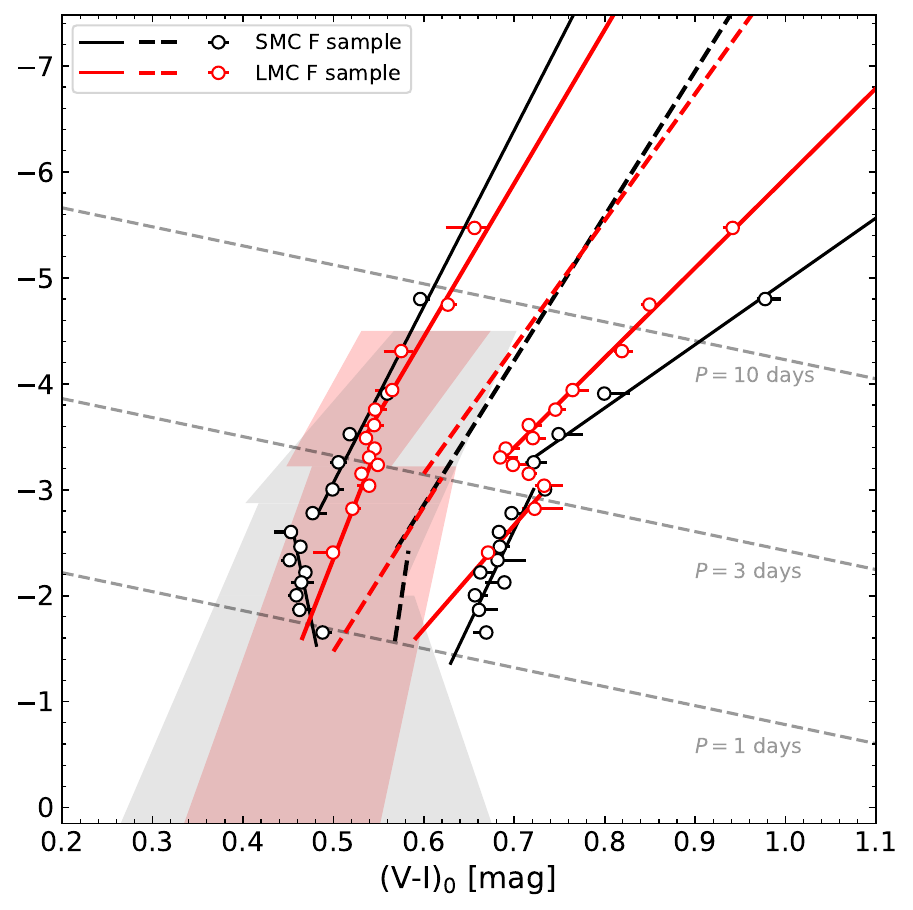}
\caption{Comparison of empirical ISs of the LMC (Paper I) and SMC (this work) in red and black, respectively. Dashed lines represent the middle of the ISs. The left panel compares only the ISs for 1O Cepheids, while the right panel shows F ISs. Shaded areas show the IS for F mode in the left panel and for 1O mode in the right panel, allowing comparison between the modes and illustrating their superposition.}
\label{fig:SMCvsLMC}
\end{figure*}

A source of uncertainty in the cosmic distance ladder that is still debated is the metallicity dependence of the P-L relations of classical Cepheids \citep[see e.g.,][]{Breuval2025,Khan2025}. Using theoretical models, it was shown that metallicity affects the shape of the IS \citep{Marconi2005}, which, in turn, may alter the P-L relations \citep{Madore1991}. To study this effect empirically, we compare the ISs for F and 1O Cepheids in the LMC and SMC (Fig.~\ref {fig:SMCvsLMC}). We included edges with breaks, in addition to the fitted median IS for both galaxies. The constant period lines shown in this figure are those computed for the SMC sample (they are slightly brighter than the same period lines calculated for the LMC). The following comparisons are based on the individual points of the IS boundaries and the mean values of the linear fits.

Taking into account the blue edge of 1O Cepheids and the red edge of F Cepheids, the IS for the LMC is narrower than for the SMC. The superposition of ISs for both pulsation modes (F and 10) is also narrower for the LMC. The breaks in slope and color of the ISs differ between the two galaxies considered, with those of the SMC occurring at shorter periods.

Regarding the IS for 1O Cepheids, the blue edge for the LMC is generally redder than the SMC, except for the longest periods. This is theoretically expected since more metal-rich pulsation models show redder IS borders. However, this explanation does not apply to the median and red edge of the empirical IS, which exhibit more complex behavior. For the LMC, both are bluer for periods longer than 3 days and shorter than 1 day. Between 1 and 2 days, the LMC median is redder, and the red edge is roughly consistent with that for the SMC. For the shortest periods, the difference between the two galaxies increases significantly, but this can be partly due to extrapolation. As mentioned in Section~\ref{subsec:xing}, this period range is poorly constrained due to low statistics of our sample below $M_I \sim -1$ mag (the faintest point for the SMC IS edge is located at $M_I=-1.5$ mag).

For F Cepheids, the ISs for both galaxies are very similar, with the LMC IS being generally narrower compared to the SMC. The blue edge of the LMC IS is redder, and its red edge is bluer. The median LMC IS is very consistent with that of the SMC. This observed similarity in the ISs of F Cepheids of different metallicities is very fortunate for their use in distance determination. It minimizes the systematic error introduced when applying empirical relations to Cepheids in different environments.
Based only on our current comparison, we can recommend using F Cepheids with periods longer than 3 days for distance determination. Given the dependence of the break position on metallicity and the lack of a similar analysis for Cepheids at solar metallicity, this period limit may be higher. Once we extend the comparison to higher-metallicity Cepheids, we will be able to provide a more accurate value.

The LMC blue edge is redder than the SMC blue edge, as expected from a theoretical perspective. However, the bluer color of the LMC red edge for F Cepheids is the opposite of expectations, and the similar color of the median ISs in both galaxies is also surprising. Further study will be needed to determine the cause of the discrepancy between theory and observation.

\begin{figure*}
\centering
\includegraphics[width=.8\hsize]{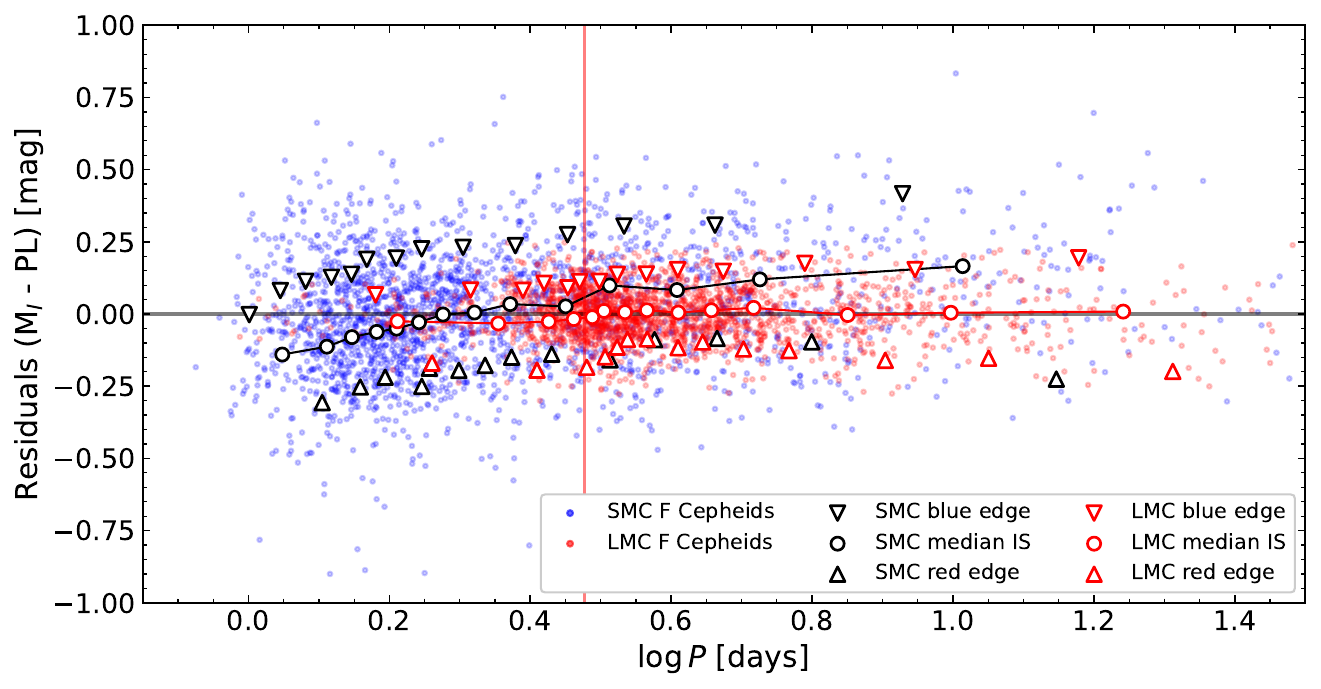}
\caption{Residuals between I-band absolute magnitudes of SMC and LMC F Cepheids and P-L relations for each galaxy while maintaining the slope fixed to the value obtained for the LMC. In addition, the residuals of the IS edges of both LMC and SMC, as well as their corresponding median parts, are shown as red and black triangles and empty circles. The SMC and LMC median ISs are connected with black and red lines, respectively. The vertical red line marks a pulsation period of 3 days.}
\label{fig:PLresiduals}
\end{figure*}

As an additional comparison, we fitted a P-L relation of the LMC and SMC samples. For the SMC, the P-L slope was fixed to the LMC value. We applied a shift to the intercept of the LMC P-L, which corresponds to the difference in distance modulus (DM) between the Magellanic clouds \cite[$\delta$ DM $= 0.5$,][]{Pietrzynski2019, Graczyk2020}, to obtain the intercept of the SMC P-L. Using the PLC relation, we converted our empirical ISs (including their medians) to the period-luminosity plane and marked them as triangles (and circles). We then computed the residuals between the absolute I-band magnitudes and the P-L relations for each galaxy. These results are shown in Fig.~\ref{fig:PLresiduals}. In this plane, it can be seen how the effect of metallicity on IS depends on pulsation period and position within the IS. It is important to note that several other physical effects, in addition to metallicity, contribute to the observed scatter in the P-L relation. These factors can include, among others, helium content, rotation, and geometrical corrections of the Magellanic Clouds \citep[see e.g.,][]{Marconi2005,Anderson2016,Breuval2022}. In Fig.~\ref{fig:PLresiduals}, it can be noted that, comparing SMC Cepheids near the red edge with LMC Cepheids near the blue edge can yield results that differ entirely from those obtained when comparing SMC Cepheids near the blue edge with LMC Cepheids near the red edge. Also, based on the median IS, the slope of the SMC relation differs from that for LMC. Although, on average, the shift between the two relations is consistent with the difference in distance moduli between the SMC and the LMC, the median values at distinct periods are considerably different. For example, using SMC Cepheids near $\log P = 0.2$ versus those near $\log P = 0.6$ would yield a systematic difference of 0.15 mag. 
In \citet{Khan2025}, the difference in P-L intercepts between the SMC and Milky Way metallicities is generally smaller than $\sim0.1$ mag (see, e.g., Fig. 9 and Table B.1 of their work), therefore using SMC Cepheid samples with different average periods may significantly affect the slope of the metallicity dependence. For the I-band, the difference between the LMC and SMC P-L intercepts in \citet{Khan2025} ranges from 0.005 to 0.075 mag, depending on the pivot period, also indicating period dependence. Moreover, generally, the wider SMC IS results in higher scatter in the P-L relation across all pulsation periods.

Taking into account the samples' locations within the instability strip may lead to a more accurate estimate of the metallicity dependence of the P–L relations. This can be done in two ways: either by selecting a sample representative of the IS part we want to study (with a proper distribution along the period and color axes), or by applying a sample-dependent correction for its position within the IS. Both procedures should allow for separating metallicity effects from other biases, homogenize results across different samples, and ultimately increase their accuracy.

\subsection{Comparison with theoretical ISs}\label{subsec:comptheory}

The theoretical IS edges have been extensively studied in the literature, exploring how various micro- and macrophysical effects alter the IS shape. Often, these studies provide the IS edges in the $\log T_{\rm eff}-\log L$ plane; thus, converting our results to this plane is necessary for comparison.

Following the same procedure as in Paper I, each point of our IS for F Cepheids in the CMD plane was converted to the $\log{T_{\rm eff}}-\log{L}$ plane using the color-temperature and bolometric correction calibration presented by \cite{Worthey2011}. The points were fitted in the same way as for the IS in the CMD. The coefficients of the empirical IS in the $\log{T_{\rm eff}}-\log{L}$ plane are presented in Table~\ref{tab:table2}. In the following part of this section, we compared our two-part IS with literature results; thus, conclusions may differ between the lower and upper parts of our IS.

\begin{figure}
\centering
\includegraphics[width=\hsize]{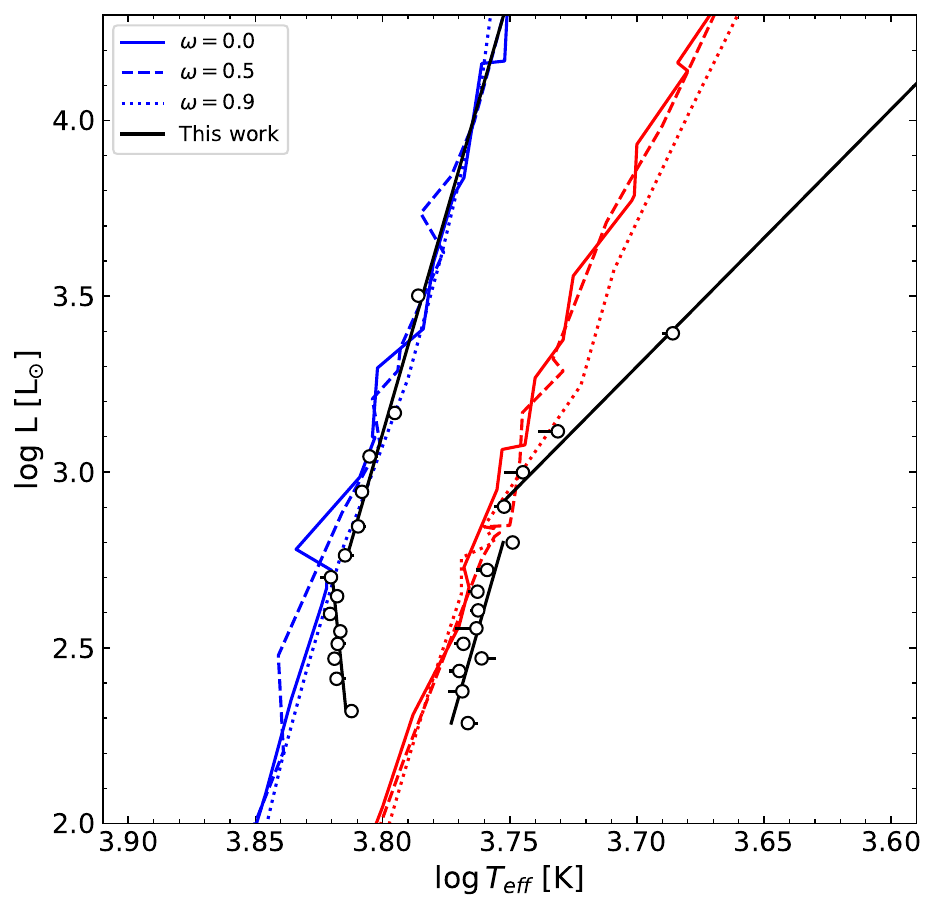}
   \caption{Comparison of theoretical F ISs presented in \cite{Anderson2016} (blue and red) and our empirical IS (black lines and black points with white centers) in an HRD. Different line styles represent different rotation rates $\omega$, expressed as a fraction of critical velocity on ZAMS.}
      \label{fig:Anderson}
\end{figure}

The theoretical ISs for F Cepheids presented by \cite{Anderson2016}, and also used in \cite{Khan2025}, are shown in Fig.~\ref{fig:Anderson}. Their results were computed using the Geneva code of stellar evolution \citep{Eggenberger2008}, for different metallicities and rotation rates, and an extended version of the method described in \citet{Saio1983} to perform a linear non-adiabatic radial pulsation analysis. We compare our results with their ISs of $Z=0.002$, and three rotation rates $\omega = 0.0$, $0.5$, and $0.9$. We reproduced their IS boundaries based on the three crossings of the IS from their Table A.3, which provides the effective temperature and luminosity of models that enter or exit the IS for each mass. Similar to the comparison performed in Paper I for the LMC, their theoretical blue edge agrees remarkably well with the bright part of our empirical one. However, the faintest portion of our blue edge is systematically cooler. Regarding their red edge, its central part is similar to our empirical red edge; however, it deviates significantly for faintest and brightest Cepheids. As for rotation, our empirical results favor models with a high rotation rate, which is consistent with the conclusions for the LMC (see Paper I).

\begin{figure*}
\centering
\includegraphics[width=\hsize]{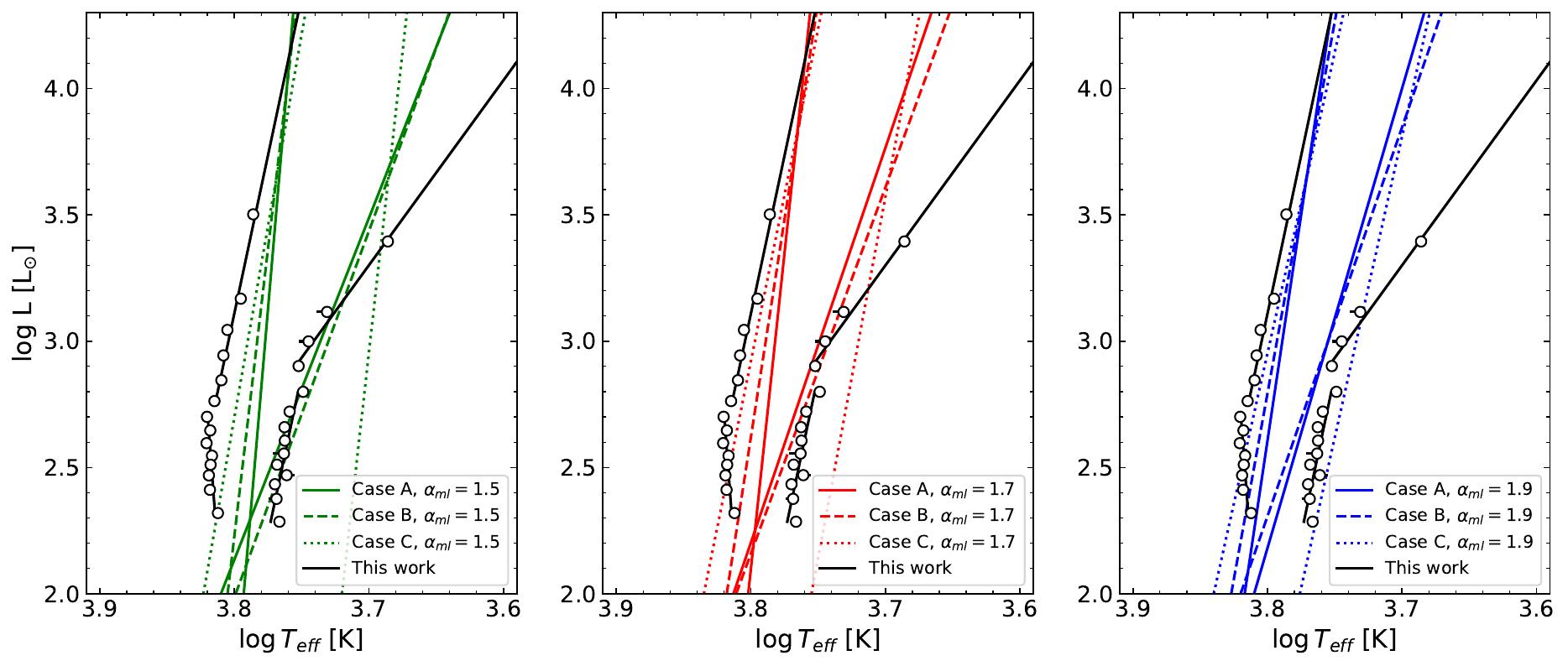}
   \caption{Comparison of theoretical F ISs for $Z = 0.004$ presented in
\citet{Somma2024} and our empirical IS (black lines and black points with white centers) in an HRD. From left to right panels, the superadiabatic convection efficiency $\alpha_{ml}$ of their models increases. In addition, different line styles represent different increases of the luminosity applied by the authors to their canonical model.  }
    \label{fig:DeSomma}
\end{figure*}

In Fig.~\ref{fig:DeSomma}, we compare our results with the theoretical IS for F Cepheids computed by \cite{Somma2022} with a metallicity of $Z=0.004$. Their nonlinear pulsational analysis was performed using the Stellingwerf code \citep{Stellingwerf1982} with updated opacity tables. Their nonlinear pulsation calculations approximate the effects of overshooting by considering an increase in the luminosity, over their canonical models (case A), of $\Delta(\log L/L_{\sun}) = 0.2$~dex (case B), and $\Delta(\log L/L_{\sun}) = 0.4$~dex (case C). Additionally, they consider three values for the superadiabatic convection efficiency of $\alpha_{ml}=1.5$, $\alpha_{ml}=1.7$, and $\alpha_{ml}=1.9$. We note good agreement for the blue edge, considering their case C, and $\alpha_{ml}=1.7$ and $\alpha_{ml}=1.9$. In the case of the red edge, our results globally align closer to their borders with the lowest $\alpha_{ml}$, and case B. Quite a good agreement is also obtained for the central part for $\alpha_{ml}=1.7$ and case B.

For theoretical ISs of \citet{Somma2022}, the agreement of the red edge with our empirical one is better than for the red edge of \cite{Anderson2016}. In particular, a $\chi^2$ calculated between the red edge considering $\alpha_{\rm ml} = 1.5$ and case B from \citet{Somma2022} and our empirical red edge, is $28\%$ lower than the one obtained considering $\omega=0.9$ from \citet{Anderson2016}. This result is expected since the location of the red edge depends on the interplay between pulsation and convection, the latter quenching pulsation, which is better described by nonlinear pulsational calculations. These nonlinear models of \citet{Somma2022} also show a gradual diminishing of the IS in the faintest part, which may support our aforementioned observation of the rapidly decreasing number of Cepheids with the shortest periods.

\begin{figure*}
\centering
\includegraphics[width=0.8\hsize]{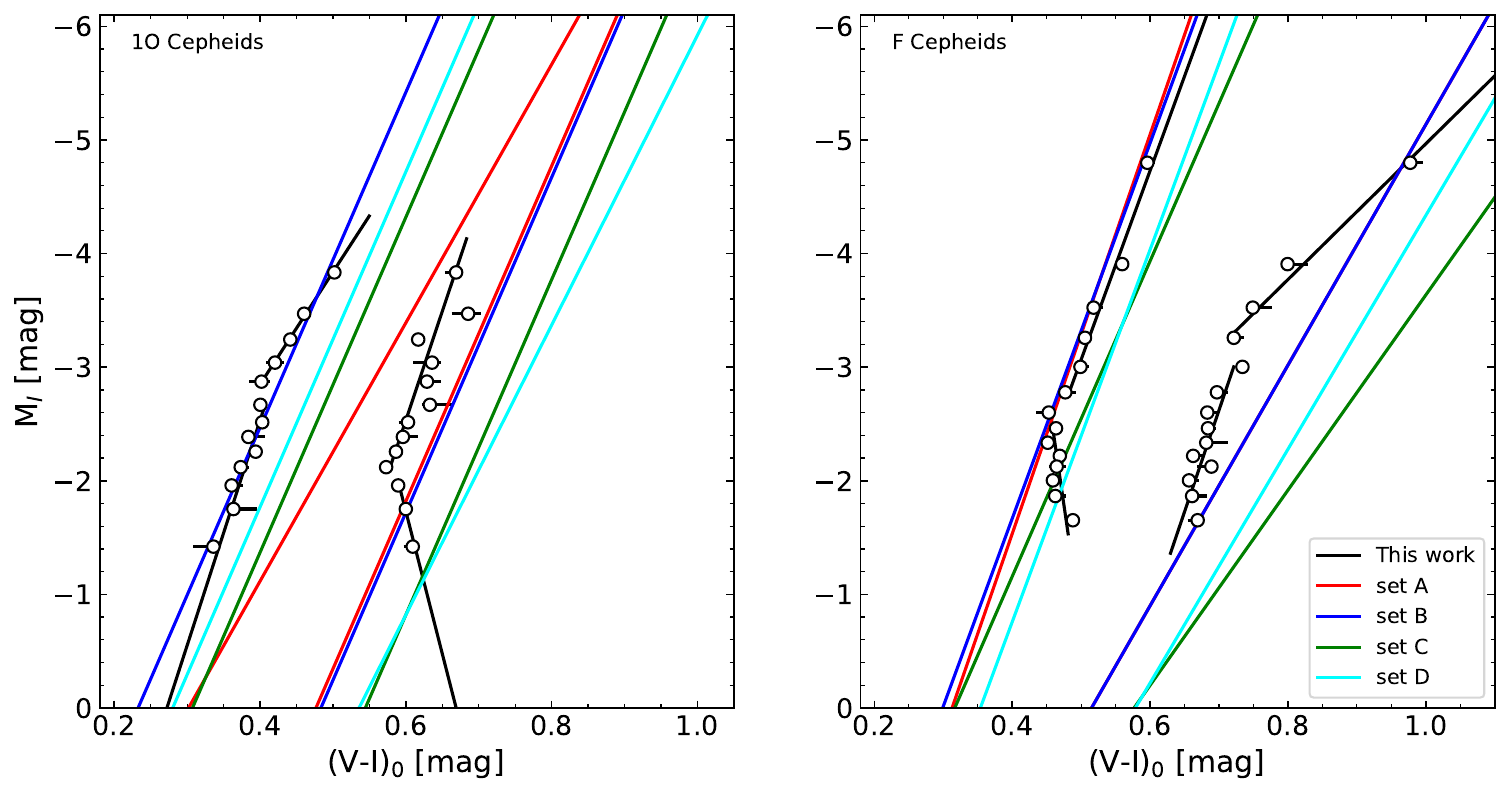}
   \caption{Comparison of theoretical ISs for $Z = 0.002$ presented in
\citet{Deka2024} and our empirical IS (black lines and black points with white centers) in a CMD. The ISs of 1O Cepheids are shown in the left panel, while ISs of F Cepheids are shown in the right panel.  Different colors represent sets of RSP convective parameters considered by the authors.}
\label{fig:Deka}
\end{figure*}

More recently, \cite{Deka2024} studied the effects on the IS using different sets of convective parameters in RSP. These parameters can be found in Table 4 of \cite{Paxton2019}, from set A to set D, increasing the complexity of the convective modeling with each consecutive set. The set A represents the simplest convective model; set B considers radiative cooling; set C takes into account turbulent pressure and turbulent flux; and set D adds all the previous effects simultaneously. A comparison between our results for 1O and F Cepheids with those obtained by \citeauthor{Deka2024} is shown in Fig.~\ref{fig:Deka}. We observed good agreement between our blue edges and their results for both samples, considering their set B. In the case of our red edge, it appears to be closer to their results using sets A and B. Contrary to the LMC case (Paper I), the RSP set D does not show satisfactory agreement with our empirical results. The latter highlights the complex interplay between metallicity and convection in the calculation of theoretical ISs.  

\section{Conclusions}\label{sec:conclusion}

We used a sample of SMC Cepheids from the OGLE-IV catalog to compute the empirical, intrinsic IS. We determined the IS edges for F and 1O Cepheids, together and separately, using I and V bands. In all cases, a change of slope of the IS was observed at pulsating periods between $1.4$ and $3$ days. The break is most prominent in the red edges, but it is also easily noticeable in the median IS.

Using a grid of evolutionary tracks computed with the stellar evolution code MESA, we investigated the connection between the observed shape of the IS and the evolution of intermediate-mass stars. Our results showed that the metallicity of the evolutionary tracks strongly influences the behavior of the blue loops. While some of this behavior coincides with observed changes in the IS (e.g., the red edge of the 1O IS), the ``oscillatory'' behavior of the blue loops does not produce observable changes in the IS' shape. This indicates that this complex feature of our models is most likely due to the physical input parameters, rather than a physical process within the star. More detailed studies of this phenomenon should be conducted to improve our understanding of Cepheids with low metallicity.

For the SMC, the lower mass limit at which the blue loops enter the IS is low enough to cover our entire Cepheid sample. This means that the first and subsequent crossing Cepheids are mixed, and depopulation cannot explain any significant change in the shape of the IS, as we proposed in Paper I for the LMC, unless we reject the models with lower metallicity. However, for both galaxies, the break also coincides with the effect of the metallicity gradient on the lower blue loop boundary. In the LMC, the transition phase between the lower and upper parts of the IS is narrow in luminosity, whereas in the SMC it is broader and can be treated as a separate component with a different slope. Therefore, the metallicity gradient can be another factor responsible for shaping the IS in a given galaxy.

We performed a more detailed comparison between our theoretical and empirical results by calculating the expected number of Cepheids as a function of period, based on the IS crossing time of our evolutionary tracks. In the case of the LMC, we found close agreement between the expected and observed numbers of Cepheids with pulsating periods longer than 3 days. On the contrary, for shorter periods (between $1$ and $3$ days), our evolutionary tracks are not able to reproduce enough second and third crossing Cepheids to match the observed number of stars. This may indicate a problem with theoretical models in this range, which do not produce sufficiently extended blue loops for lower mass stars. For periods shorter than $1$ day, the expected number of stars is larger than the observed one. On the other hand, the SMC results showed close agreement between empirical and theoretical results for pulsating periods as short as $1$ day. For shorter periods, the expected number of stars is much larger than the observed number, which is likely related to a problem in our linear pulsating models and the extrapolation of our empirical IS where the statistics are low. Generally, the results for both galaxies indicate that short-period Cepheids remain a modeling challenge for current stellar evolution codes. A similar conclusion, although for other reasons, was drawn in our recent study of a short-period binary Cepheid OGLE LMC-CEP-1347 \citep{Espinoza2025}. The analysis of this and other binary Cepheids also indicates that the number of short-period Cepheids can be affected by merger origin from lower mass stars \citep{Pilecki2022, Pilecki2024a}.

The elongated geometry of the SMC poses an additional challenge for the detailed study of the IS, as it increases the uncertainty in the calculated distances to Cepheids. To study its impact on the shape of the IS, we applied our method to a subsample of Cepheids near the center of the SMC. We report a significant shift of the blue edge of the IS to cooler temperatures. This effect is also observed at the red edge, but to a lesser extent. A possible explanation is that this subsample of Cepheids shows higher reddening than indicated by the reddening maps. Additionally, a shift to redder colors can be explained by a higher metallicity than in the full sample. The latter is consistent with previous studies, although the magnitude of this effect remains uncertain.

To test the effect of metallicity on the IS, we compared our empirical results for the SMC and LMC. The SMC showed a wider IS for both pulsating modes. Furthermore, the blue edges in both galaxies follow the expected theoretical trend: the more metal-rich LMC blue edge is redder than its SMC counterpart. The red edge, on the other hand, presents a more complex behavior. Its position is strongly influenced by changes in slope at different luminosities, and it displays a metallicity dependence opposite to theoretical predictions. This implies that accurately locating a Cepheid sample within the IS is crucial for precisely determining the metallicity dependence in P-L relations. In particular, selecting Cepheids located in different parts of the IS could yield different determined metallicity effects. On the other hand, the transition of 1O to F Cepheids, delimited by the red edge of the 1O IS and the blue edge of the F IS, does present a complex dependence on metallicity. This could be further explored by studying the precise location of double-mode Cepheids in a CMD, which may be switching between F and 1O pulsation modes. An important conclusion from the comparison of ISs for the LMC and SMC is that, within this metallicity range, F Cepheids with periods longer than 3 days should be relatively safe for distance determination. However, at higher metallicities, there is no similar analysis. In the future, we will extend our study to higher metallicities (e.g., in the Milky Way) and provide a precise recommendation considering this environment.

We compared our empirically determined IS with prior theoretical studies. The bright part of the F blue edge of \cite{Anderson2016} shows good agreement with our empirical IS, while the bright end of their F red edge deviates from it significantly. The edges obtained in \cite{Somma2022} can reproduce our IS well, but no unique combination of their parameters can reproduce both red and blue edges. Our blue edge is closer to their Case C, considering moderate and high convection efficiency ($\alpha_{ml}=1.7$ and $\alpha_{ml}=1.9$). On the other hand, our red edge agrees better with their results for Case B and low convection efficiency ($\alpha_{ml}=1.5$). In the case of the results by \cite{Deka2024}, only our blue edge can be reproduced satisfactorily, in particular using their set B (for both 1O and F Cepheids), or their set A (only F Cepheids). The slopes of our red edge cannot be reproduced by their results. 

Our study of the instability strip in the Magellanic Clouds demonstrates its value for future Cepheid research. This includes, among others, constraining theoretical evolutionary and pulsational models, investigating the behavior of short-period Cepheids, and understanding the role of metallicity in shaping the relation between the period, luminosity, and color of Cepheids. Furthermore, these IS edges can be applied to other low-metallicity environments with fewer stars to efficiently identify and select Cepheids, also depending on their pulsation mode.

\begin{acknowledgements}
We thank the anonymous referee for the constructive comments and suggestion. We acknowledge support from the Polish National Science Center grant SONATA BIS 2020/38/E/ST9/00486. This research has made use of NASA's Astrophysics Data System Service. 
\end{acknowledgements}

\bibliographystyle{aa.bst}
\bibliography{main}

@ARTICLE{Skowron2021,
       author = {{Skowron}, D.~M. and {Skowron}, J. and {Udalski}, A. and {Szyma{\'n}ski}, M.~K. and {Soszy{\'n}ski}, I. and {Wyrzykowski}, {\L}. and {Ulaczyk}, K. and {Poleski}, R. and {Koz{\l}owski}, S. and {Pietrukowicz}, P. and {Mr{\'o}z}, P. and {Rybicki}, K. and {Iwanek}, P. and {Wrona}, M. and {Gromadzki}, M.},
        title = "{OGLE-ing the Magellanic System: Optical Reddening Maps of the Large and Small Magellanic Clouds from Red Clump Stars}",
      journal = {\apjs},
         year = 2021,
        month = feb,
       volume = {252},
       number = {2},
          eid = {23},
        pages = {23},
          doi = {10.3847/1538-4365/abcb81},
archivePrefix = {arXiv},
       eprint = {2006.02448},
 primaryClass = {astro-ph.SR},
       adsurl = {https://ui.adsabs.harvard.edu/abs/2021ApJS..252...23S}
}

@ARTICLE{Jacyszyn2016,
       author = {{Jacyszyn-Dobrzeniecka}, A.~M. and {Skowron}, D.~M. and {Mr{\'o}z}, P. and {Skowron}, J. and {Soszy{\'n}ski}, I. and {Udalski}, A. and {Pietrukowicz}, P. and {Koz{\l}owski}, S. and {Wyrzykowski}, {\L}. and {Poleski}, R. and {Pawlak}, M. and {Szyma{\'n}ski}, M.~K. and {Ulaczyk}, K.},
        title = "{OGLE-ing the Magellanic System: Three-Dimensional Structure of the Clouds and the Bridge Using Classical Cepheids}",
      journal = {\actaa},
         year = 2016,
        month = jun,
       volume = {66},
       number = {2},
        pages = {149-196},
archivePrefix = {arXiv},
       eprint = {1602.09141},
 primaryClass = {astro-ph.GA},
       adsurl = {https://ui.adsabs.harvard.edu/abs/2016AcA....66..149J}
}

@ARTICLE{2017AcA....67..103S,
       author = {{Soszy{\'n}ski}, I. and {Udalski}, A. and {Szyma{\'n}ski}, M.~K. and {Wyrzykowski}, {\L}. and {Ulaczyk}, K. and {Poleski}, R. and {Pietrukowicz}, P. and {Koz{\l}owski}, S. and {Skowron}, D.~M. and {Skowron}, J. and {Mr{\'o}z}, P. and {Pawlak}, M.},
        title = "{Concluding Henrietta Leavitt's Work on Classical Cepheids in the Magellanic System and Other Updates of the OGLE Collection of Variable Stars}",
      journal = {\actaa},
         year = 2017,
        month = jun,
       volume = {67},
       number = {2},
        pages = {103-113},
          doi = {10.32023/0001-5237/67.2.1},
archivePrefix = {arXiv},
       eprint = {1706.09452},
 primaryClass = {astro-ph.SR},
       adsurl = {https://ui.adsabs.harvard.edu/abs/2017AcA....67..103S}
}

@ARTICLE{Paxton2011,
  author = {{Paxton}, B. and {Bildsten}, L. and {Dotter}, A. and {Herwig}, F. and {Lesaffre}, P. and {Timmes}, F.},
  title = {{Modules for Experiments in Stellar Astrophysics (MESA)}},
  journal = {\apjs},
  archivePrefix = {arXiv},
  eprint = {1009.1622},
  primaryClass = {astro-ph.SR},
  year = {2011},
  month = {jan},
  volume = {192},
  eid = {3},
  pages = {3},
  doi = {10.1088/0067-0049/192/1/3},
  adsurl = {https://ui.adsabs.harvard.edu/abs/2011ApJS..192....3P},
}

@ARTICLE{Paxton2013,
  author = {{Paxton}, B. and {Cantiello}, M. and {Arras}, P. and {Bildsten}, L. and {Brown}, E.~F. and {Dotter}, A. and {Mankovich}, C. and {Montgomery}, M.~H. and {Stello}, D. and {Timmes}, F.~X. and {Townsend}, R.},
  title = {{Modules for Experiments in Stellar Astrophysics (MESA): Planets, Oscillations, Rotation, and Massive Stars}},
  journal = {\apjs},
  archivePrefix = {arXiv},
  eprint = {1301.0319},
  primaryClass = {astro-ph.SR},
  year = {2013},
  month = {sep},
  volume = {208},
  eid = {4},
  pages = {4},
  doi = {10.1088/0067-0049/208/1/4},
  adsurl = {https://ui.adsabs.harvard.edu/abs/2013ApJS..208....4P},
}

@ARTICLE{Paxton2015,
  author = {{Paxton}, B. and {Marchant}, P. and {Schwab}, J. and {Bauer}, E.~B. and {Bildsten}, L. and {Cantiello}, M. and {Dessart}, L. and {Farmer}, R. and {Hu}, H. and {Langer}, N. and {Townsend}, R.~H.~D. and {Townsley}, D.~M. and {Timmes}, F.~X.},
  title = {{Modules for Experiments in Stellar Astrophysics (MESA): Binaries, Pulsations, and Explosions}},
  journal = {\apjs},
  archivePrefix = {arXiv},
  eprint = {1506.03146},
  primaryClass = {astro-ph.SR},
  year = {2015},
  month = {sep},
  volume = {220},
  eid = {15},
  pages = {15},
  doi = {10.1088/0067-0049/220/1/15},
  adsurl = {https://ui.adsabs.harvard.edu/abs/2015ApJS..220...15P},
}

@ARTICLE{Paxton2018,
  author = {{Paxton}, B. and {Schwab}, J. and {Bauer}, E.~B. and {Bildsten}, L. and {Blinnikov}, S. and {Duffell}, P. and {Farmer}, R. and {Goldberg}, J.~A. and {Marchant}, P. and {Sorokina}, E. and {Thoul}, A. and {Townsend}, R.~H.~D. and {Timmes}, F.~X.},
  title = {{Modules for Experiments in Stellar Astrophysics (MESA): Convective Boundaries, Element Diffusion, and Massive Star Explosions}},
  journal = {\apjs},
  archivePrefix = {arXiv},
  eprint = {1710.08424},
  primaryClass = {astro-ph.SR},
  year = {2018},
  month = {feb},
  volume = {234},
  eid = {34},
  pages = {34},
  doi = {10.3847/1538-4365/aaa5a8},
  adsurl = {https://ui.adsabs.harvard.edu/abs/2018ApJS..234...34P},
}

@ARTICLE{Paxton2019,
       author = {{Paxton}, Bill and {Smolec}, R. and {Schwab}, Josiah and {Gautschy}, A. and
         {Bildsten}, Lars and {Cantiello}, Matteo and {Dotter}, Aaron and
         {Farmer}, R. and {Goldberg}, Jared A. and {Jermyn}, Adam S. and
         {Kanbur}, S.~M. and {Marchant}, Pablo and {Thoul}, Anne and
         {Townsend}, Richard H.~D. and {Wolf}, William M. and {Zhang}, Michael and
         {Timmes}, F.~X.},
        title = "{Modules for Experiments in Stellar Astrophysics (MESA): Pulsating Variable Stars, Rotation, Convective Boundaries, and Energy Conservation}",
      journal = {\apjs},
         year = "2019",
        month = "Jul",
       volume = {243},
       number = {1},
          eid = {10},
        pages = {10},
          doi = {10.3847/1538-4365/ab2241},
archivePrefix = {arXiv},
       eprint = {1903.01426},
 primaryClass = {astro-ph.SR},
       adsurl = {https://ui.adsabs.harvard.edu/abs/2019ApJS..243...10P}
}

@ARTICLE{Bauer1999,
       author = {{Bauer}, F. and {Afonso}, C. and {Albert}, J.~N. and {Alard}, C. and {Andersen}, J. and {Ansari}, R. and {Aubourg}, {\'E}. and {Bareyre}, P. and {Beaulieu}, J.~P. and {Bouquet}, A. and {Char}, S. and {Charlot}, X. and {Couchot}, F. and {Coutures}, C. and {Derue}, F. and {Ferlet}, R. and {Gaucherel}, C. and {Glicenstein}, J.~F. and {Goldman}, B. and {Gould}, A. and {Graff}, D. and {Gros}, M. and {Haissinski}, J. and {Hamilton}, J.~C. and {Hardin}, D. and {de Kat}, J. and {Kim}, A. and {Lasserre}, T. and {Lesquoy}, {\'E}. and {Loup}, C. and {Magneville}, C. and {Mansoux}, B. and {Marquette}, J.~B. and {Maurice}, {\'E}. and {Milsztajn}, A. and {Moniez}, M. and {Palanque-Delabrouille}, N. and {Perdereau}, O. and {Pr{\'e}vot}, L. and {Renault}, C. and {Regnault}, N. and {Rich}, J. and {Spiro}, M. and {Vidal-Madjar}, A. and {Vigroux}, L. and {Zylberajch}, S.},
        title = "{A slope variation in the period-luminosity relation for short period SMC Cepheids}",
      journal = {\aap},
         year = 1999,
        month = aug,
       volume = {348},
        pages = {175-183},
       adsurl = {https://ui.adsabs.harvard.edu/abs/1999A&A...348..175E}
}

@ARTICLE{Sharpee2002,
       author = {{Sharpee}, Brian and {Stark}, Michele and {Pritzl}, Barton and {Smith}, Horace and {Silbermann}, Nancy and {Wilhelm}, Ronald and {Walker}, Alistair},
        title = "{BV Photometry of Variable Stars in the Northeast Arm of the Small Magellanic Cloud}",
      journal = {\aj},
         year = 2002,
        month = jun,
       volume = {123},
       number = {6},
        pages = {3216-3254},
          doi = {10.1086/340185},
archivePrefix = {arXiv},
       eprint = {astro-ph/0202259},
 primaryClass = {astro-ph},
       adsurl = {https://ui.adsabs.harvard.edu/abs/2002AJ....123.3216S}
}

@ARTICLE{Sandage2004,
       author = {{Sandage}, A. and {Tammann}, G.~A. and {Reindl}, B.},
        title = "{New period-luminosity and period-color relations of classical Cepheids. II. Cepheids in LMC}",
      journal = {\aap},
         year = 2004,
        month = sep,
       volume = {424},
        pages = {43-71},
          doi = {10.1051/0004-6361:20040222},
archivePrefix = {arXiv},
       eprint = {astro-ph/0402424},
 primaryClass = {astro-ph},
       adsurl = {https://ui.adsabs.harvard.edu/abs/2004A&A...424...43S}
}

@ARTICLE{Sandage2009,
       author = {{Sandage}, A. and {Tammann}, G.~A. and {Reindl}, B.},
        title = "{New period-luminosity and period-color relations of classical Cepheids. III. Cepheids in SMC}",
      journal = {\aap},
         year = 2009,
        month = jan,
       volume = {493},
       number = {2},
        pages = {471-479},
          doi = {10.1051/0004-6361:200810550},
archivePrefix = {arXiv},
       eprint = {0810.1780},
 primaryClass = {astro-ph},
       adsurl = {https://ui.adsabs.harvard.edu/abs/2009A&A...493..471S}
}

@ARTICLE{Smolec2008,
       author = {{Smolec}, R. and {Moskalik}, P.},
        title = "{Convective Hydrocodes for Radial Stellar Pulsation. Physical and Numerical Formulation}",
      journal = {\actaa},
         year = 2008,
        month = sep,
       volume = {58},
        pages = {193-232},
archivePrefix = {arXiv},
       eprint = {0809.1979},
 primaryClass = {astro-ph},
       adsurl = {https://ui.adsabs.harvard.edu/abs/2008AcA....58..193S}
}

@ARTICLE{Anderson2016,
       author = {{Anderson}, R.~I. and {Saio}, H. and {Ekstr{\"o}m}, S. and {Georgy}, C. and {Meynet}, G.},
        title = "{On the effect of rotation on populations of classical Cepheids. II. Pulsation analysis for metallicities 0.014, 0.006, and 0.002}",
      journal = {\aap},
         year = 2016,
        month = jun,
       volume = {591},
          eid = {A8},
        pages = {A8},
          doi = {10.1051/0004-6361/201528031},
archivePrefix = {arXiv},
       eprint = {1604.05691},
 primaryClass = {astro-ph.SR},
       adsurl = {https://ui.adsabs.harvard.edu/abs/2016A&A...591A...8A}
}

@INPROCEEDINGS{PelLub1978,
       author = {{Pel}, J.~W. and {Lub}, J.},
        title = "{Physical Properties of Cepheids and RR Lyrae Stars}",
    booktitle = {The HR Diagram - The 100th Anniversary of Henry Norris Russell},
         year = 1978,
       editor = {{Philip}, A.~G. Davis and {Hayes}, D.~S.},
       volume = {80},
        month = jan,
        pages = {229},
       adsurl = {https://ui.adsabs.harvard.edu/abs/1978IAUS...80..229P}
}

@ARTICLE{Fernie1990,
       author = {{Fernie}, J.~D.},
        title = "{The Structure of the Cepheid Instability Strip}",
      journal = {\apj},
         year = 1990,
        month = may,
       volume = {354},
        pages = {295},
          doi = {10.1086/168689},
       adsurl = {https://ui.adsabs.harvard.edu/abs/1990ApJ...354..295F}
}

@ARTICLE{Somma2022,
       author = {{De Somma}, Giulia and {Marconi}, Marcella and {Molinaro}, Roberto and {Ripepi}, Vincenzo and {Leccia}, Silvio and {Musella}, Ilaria},
        title = "{An Updated Metal-dependent Theoretical Scenario for Classical Cepheids}",
      journal = {\apjs},
         year = 2022,
        month = sep,
       volume = {262},
       number = {1},
          eid = {25},
        pages = {25},
          doi = {10.3847/1538-4365/ac7f3b},
archivePrefix = {arXiv},
       eprint = {2206.11154},
 primaryClass = {astro-ph.SR},
       adsurl = {https://ui.adsabs.harvard.edu/abs/2022ApJS..262...25D}
}

@ARTICLE{Turner2001,
       author = {{Turner}, D.~G.},
        title = "{A New Mapping of the Cepheid Instability Strip}",
      journal = {Odessa Astronomical Publications},
         year = 2001,
        month = dec,
       volume = {14},
        pages = {166-169},
       adsurl = {https://ui.adsabs.harvard.edu/abs/2001OAP....14..166T}
}

@ARTICLE{Tammann2003,
       author = {{Tammann}, G.~A. and {Sandage}, A. and {Reindl}, B.},
        title = "{New Period-Luminosity and Period-Color relations of classical Cepheids: I. Cepheids in the Galaxy}",
      journal = {\aap},
         year = 2003,
        month = jun,
       volume = {404},
        pages = {423-448},
          doi = {10.1051/0004-6361:20030354},
archivePrefix = {arXiv},
       eprint = {astro-ph/0303378},
 primaryClass = {astro-ph},
       adsurl = {https://ui.adsabs.harvard.edu/abs/2003A&A...404..423T}
}

@ARTICLE{Pietrzynski2019,
       author = {{Pietrzy{\'n}ski}, G. and {Graczyk}, D. and {Gallenne}, A. and {Gieren}, W. and {Thompson}, I.~B. and {Pilecki}, B. and {Karczmarek}, P. and {G{\'o}rski}, M. and {Suchomska}, K. and {Taormina}, M. and {Zgirski}, B. and {Wielg{\'o}rski}, P. and {Ko{\l}aczkowski}, Z. and {Konorski}, P. and {Villanova}, S. and {Nardetto}, N. and {Kervella}, P. and {Bresolin}, F. and {Kudritzki}, R.~P. and {Storm}, J. and {Smolec}, R. and {Narloch}, W.},
        title = "{A distance to the Large Magellanic Cloud that is precise to one per cent}",
      journal = {\nat},
         year = 2019,
        month = mar,
       volume = {567},
       number = {7747},
        pages = {200-203},
          doi = {10.1038/s41586-019-0999-4},
archivePrefix = {arXiv},
       eprint = {1903.08096},
 primaryClass = {astro-ph.GA},
       adsurl = {https://ui.adsabs.harvard.edu/abs/2019Natur.567..200P}
}

@ARTICLE{Hidalgo2018,
       author = {{Hidalgo}, Sebastian L. and {Pietrinferni}, Adriano and {Cassisi}, Santi and {Salaris}, Maurizio and {Mucciarelli}, Alessio and {Savino}, Alessandro and {Aparicio}, Antonio and {Silva Aguirre}, Victor and {Verma}, Kuldeep},
        title = "{The Updated BaSTI Stellar Evolution Models and Isochrones. I. Solar-scaled Calculations}",
      journal = {\apj},
         year = 2018,
        month = apr,
       volume = {856},
       number = {2},
          eid = {125},
        pages = {125},
          doi = {10.3847/1538-4357/aab158},
archivePrefix = {arXiv},
       eprint = {1802.07319},
 primaryClass = {astro-ph.GA},
       adsurl = {https://ui.adsabs.harvard.edu/abs/2018ApJ...856..125H}
}

@ARTICLE{Eggenberger2008,
       author = {{Eggenberger}, P. and {Meynet}, G. and {Maeder}, A. and {Hirschi}, R. and
         {Charbonnel}, C. and {Talon}, S. and {Ekstr{\"o}m}, S.},
        title = "{The Geneva stellar evolution code}",
      journal = {\apss},
         year = 2008,
        month = aug,
       volume = {316},
       number = {1-4},
        pages = {43-54},
          doi = {10.1007/s10509-007-9511-y},
       adsurl = {https://ui.adsabs.harvard.edu/abs/2008Ap&SS.316...43E}
}

@ARTICLE{2004Xua,
       author = {{Xu}, H.~Y. and {Li}, Y.},
        title = "{Blue loops of intermediate mass stars . I. CNO cycles and blue loops}",
      journal = {\aap},
         year = 2004,
        month = apr,
       volume = {418},
        pages = {213-224},
          doi = {10.1051/0004-6361:20040024},
       adsurl = {https://ui.adsabs.harvard.edu/abs/2004A&A...418..213X}
}

@ARTICLE{Walmswell2015,
       author = {{Walmswell}, J.~J. and {Tout}, C.~A. and {Eldridge}, J.~J.},
        title = "{On the blue loops of intermediate-mass stars}",
      journal = {\mnras},
         year = 2015,
        month = mar,
       volume = {447},
       number = {3},
        pages = {2951-2960},
          doi = {10.1093/mnras/stu2666},
archivePrefix = {arXiv},
       eprint = {1502.04311},
 primaryClass = {astro-ph.SR},
       adsurl = {https://ui.adsabs.harvard.edu/abs/2015MNRAS.447.2951W}
}

@article{Worthey2011,
doi = {10.1088/0067-0049/193/1/1},
url = {https://dx.doi.org/10.1088/0067-0049/193/1/1},
year = {2011},
month = {jan},
publisher = {The American Astronomical Society},
volume = {193},
number = {1},
pages = {1},
author = {{Worthey}, G and {Lee}, H},
title = {AN EMPIRICAL UBV RI JHK COLOR–TEMPERATURE CALIBRATION FOR STARS},
journal = {The Astrophysical Journal Supplement Series}
}

@ARTICLE{Jermyn2023,
       author = {{Jermyn}, Adam S. and {Bauer}, Evan B. and {Schwab}, Josiah and {Farmer}, R. and {Ball}, Warrick H. and {Bellinger}, Earl P. and {Dotter}, Aaron and {Joyce}, Meridith and {Marchant}, Pablo and {Mombarg}, Joey S.~G. and {Wolf}, William M. and {Sunny Wong}, Tin Long and {Cinquegrana}, Giulia C. and {Farrell}, Eoin and {Smolec}, R. and {Thoul}, Anne and {Cantiello}, Matteo and {Herwig}, Falk and {Toloza}, Odette and {Bildsten}, Lars and {Townsend}, Richard H.~D. and {Timmes}, F.~X.},
        title = "{Modules for Experiments in Stellar Astrophysics (MESA): Time-dependent Convection, Energy Conservation, Automatic Differentiation, and Infrastructure}",
      journal = {\apjs},
         year = 2023,
        month = mar,
       volume = {265},
       number = {1},
          eid = {15},
        pages = {15},
          doi = {10.3847/1538-4365/acae8d},
archivePrefix = {arXiv},
       eprint = {2208.03651},
 primaryClass = {astro-ph.SR},
       adsurl = {https://ui.adsabs.harvard.edu/abs/2023ApJS..265...15J}
}

@ARTICLE{Grevesse1998,
       author = {{Grevesse}, N. and {Sauval}, A.~J.},
        title = "{Standard Solar Composition}",
      journal = {\ssr},
         year = 1998,
        month = may,
       volume = {85},
        pages = {161-174},
          doi = {10.1023/A:1005161325181},
       adsurl = {https://ui.adsabs.harvard.edu/abs/1998SSRv...85..161G}
}

@ARTICLE{Reimers1975,
       author = {{Reimers}, D.},
        title = "{Circumstellar absorption lines and mass loss from red giants.}",
      journal = {Memoires of the Societe Royale des Sciences de Liege},
         year = 1975,
        month = jan,
       volume = {8},
        pages = {369-382},
       adsurl = {https://ui.adsabs.harvard.edu/abs/1975MSRSL...8..369R}
}

@BOOK{catelan2015,
       author = {{Catelan}, M. and {Smith}, H.~A.},
        title = "{Pulsating Stars}",
         year = 2015,
       adsurl = {https://ui.adsabs.harvard.edu/abs/2015pust.book.....C},
      publisher={Wiley-VCH, Weinheim}
}

@ARTICLE{Schlegel1998,
       author = {{Schlegel}, David J. and {Finkbeiner}, Douglas P. and {Davis}, Marc},
        title = "{Maps of Dust Infrared Emission for Use in Estimation of Reddening and Cosmic Microwave Background Radiation Foregrounds}",
      journal = {\apj},
         year = 1998,
        month = jun,
       volume = {500},
       number = {2},
        pages = {525-553},
          doi = {10.1086/305772},
archivePrefix = {arXiv},
       eprint = {astro-ph/9710327},
 primaryClass = {astro-ph},
       adsurl = {https://ui.adsabs.harvard.edu/abs/1998ApJ...500..525S}
}

@ARTICLE{Zhao2023,
       author = {{Zhao}, Liuyan and {Song}, Hanfeng and {Meynet}, Georges and {Maeder}, Andre and {Ekstr{\"o}m}, Sylvia and {Zhang}, Ruiyu and {Qin}, Ying and {Qi}, Shitao and {Zhan}, Qiong},
        title = "{The evolutionary properties of the blue loop under the influence of rapid rotation and low metallicity}",
      journal = {\aap},
         year = 2023,
        month = jun,
       volume = {674},
          eid = {A92},
        pages = {A92},
          doi = {10.1051/0004-6361/202245665},
       adsurl = {https://ui.adsabs.harvard.edu/abs/2023A&A...674A..92Z}
}

@ARTICLE{Espinoza2024,
       author = {{Espinoza-Arancibia}, F. and {Pilecki}, B. and {Pietrzy{\'n}ski}, G. and {Smolec}, R. and {Kervella}, P.},
        title = "{Empirical instability strip for classical Cepheids. I. The Large Magellanic Cloud galaxy}",
      journal = {\aap},
         year = 2024,
        month = feb,
       volume = {682},
          eid = {A185},
        pages = {A185},
          doi = {10.1051/0004-6361/202347804},
archivePrefix = {arXiv},
       eprint = {2311.15849},
 primaryClass = {astro-ph.SR},
       adsurl = {https://ui.adsabs.harvard.edu/abs/2024A&A...682A.185E}
}

@ARTICLE{Graczyk2020,
       author = {{Graczyk}, Dariusz and {Pietrzy{\'n}ski}, Grzegorz and {Thompson}, Ian B. and {Gieren}, Wolfgang and {Zgirski}, Bart{\l}omiej and {Villanova}, Sandro and {G{\'o}rski}, Marek and {Wielg{\'o}rski}, Piotr and {Karczmarek}, Paulina and {Narloch}, Weronika and {Pilecki}, Bogumi{\l} and {Taormina}, Monica and {Smolec}, Rados{\l}aw and {Suchomska}, Ksenia and {Gallenne}, Alexandre and {Nardetto}, Nicolas and {Storm}, Jesper and {Kudritzki}, Rolf-Peter and {Ka{\l}uszy{\'n}ski}, Miko{\l}aj and {Pych}, Wojciech},
        title = "{A Distance Determination to the Small Magellanic Cloud with an Accuracy of Better than Two Percent Based on Late-type Eclipsing Binary Stars}",
      journal = {\apj},
         year = 2020,
        month = nov,
       volume = {904},
       number = {1},
          eid = {13},
        pages = {13},
          doi = {10.3847/1538-4357/abbb2b},
archivePrefix = {arXiv},
       eprint = {2010.08754},
 primaryClass = {astro-ph.GA},
       adsurl = {https://ui.adsabs.harvard.edu/abs/2020ApJ...904...13G}
}

@ARTICLE{Deka2024,
       author = {{Deka}, Mami and {Bellinger}, Earl P. and {Kanbur}, Shashi M. and {Deb}, Sukanta and {Bhardwaj}, Anupam and {Randall}, Hugh Riley and {Kalici}, Selim and {Das}, Susmita},
        title = "{Bridging theory and observations in stellar pulsations: the impact of convection and metallicity on the instability strips of classical and type-II cepheids}",
      journal = {\mnras},
         year = 2024,
        month = jun,
       volume = {530},
       number = {4},
        pages = {5099-5119},
          doi = {10.1093/mnras/stae1136},
archivePrefix = {arXiv},
       eprint = {2404.17141},
 primaryClass = {astro-ph.SR},
       adsurl = {https://ui.adsabs.harvard.edu/abs/2024MNRAS.530.5099D}
}

@ARTICLE{Madore1991,
       author = {{Madore}, Barry F. and {Freedman}, Wendy L.},
        title = "{The Cepheid Distance Scale}",
      journal = {\pasp},
         year = 1991,
        month = sep,
       volume = {103},
        pages = {933},
          doi = {10.1086/132911},
       adsurl = {https://ui.adsabs.harvard.edu/abs/1991PASP..103..933M}
}

@ARTICLE{Choudhury2018,
       author = {{Choudhury}, S. and {Subramaniam}, A. and {Cole}, A.~A. and {Sohn}, Y. -J.},
        title = "{Photometric metallicity map of the Small Magellanic Cloud}",
      journal = {\mnras},
         year = 2018,
        month = apr,
       volume = {475},
       number = {4},
        pages = {4279-4297},
          doi = {10.1093/mnras/sty087},
archivePrefix = {arXiv},
       eprint = {1801.03403},
 primaryClass = {astro-ph.GA},
       adsurl = {https://ui.adsabs.harvard.edu/abs/2018MNRAS.475.4279C}
}

@INPROCEEDINGS{Smolec2023,
       author = {{Smolec}, Radoslaw and {Ziolkowska}, Oliwia and {Rathour}, Rajeev Singh and {Hocde}, Vincent},
        title = "{Stellar evolutionary tracks for medium mass stars - effects of microphysics, core and envelope overshooting and mass loss}",
    booktitle = {PLATO Stellar Science Conference 2023},
         year = 2023,
        month = jul,
          eid = {7},
        pages = {7},
          doi = {10.5281/zenodo.8207361},
       adsurl = {https://ui.adsabs.harvard.edu/abs/2023plat.confE...7S}
}

@ARTICLE{Breuval2024,
       author = {{Breuval}, Louise and {Riess}, Adam G. and {Casertano}, Stefano and {Yuan}, Wenlong and {Macri}, Lucas M. and {Romaniello}, Martino and {Murakami}, Yukei S. and {Scolnic}, Daniel and {Anand}, Gagandeep S. and {Soszy{\'n}ski}, Igor},
        title = "{Small Magellanic Cloud Cepheids Observed with the Hubble Space Telescope Provide a New Anchor for the SH0ES Distance Ladder}",
      journal = {\apj},
         year = 2024,
        month = sep,
       volume = {973},
       number = {1},
          eid = {30},
        pages = {30},
          doi = {10.3847/1538-4357/ad630e},
archivePrefix = {arXiv},
       eprint = {2404.08038},
 primaryClass = {astro-ph.CO},
       adsurl = {https://ui.adsabs.harvard.edu/abs/2024ApJ...973...30B}
}

@ARTICLE{Murray2024,
       author = {{Murray}, Claire E. and {Hasselquist}, Sten and {Peek}, Joshua E.~G. and {Lindberg}, Christina Willecke and {Almeida}, Andres and {Choi}, Yumi and {Craig}, Jessica E.~M. and {D{\'e}nes}, Helga and {Dickey}, John M. and {Di Teodoro}, Enrico M. and {Federrath}, Christoph and {Gerrard}, Isabella. A. and {Gibson}, Steven J. and {Leahy}, Denis and {Lee}, Min-Young and {Lynn}, Callum and {Ma}, Yik Ki and {Marchal}, Antoine and {McClure-Griffiths}, N.~M. and {Nidever}, David and {Nguyen}, Hiep and {Pingel}, Nickolas M. and {Tarantino}, Elizabeth and {Uscanga}, Lucero and {van Loon}, Jacco Th.},
        title = "{A Galactic Eclipse: The Small Magellanic Cloud Is Forming Stars in Two Superimposed Systems}",
      journal = {\apj},
         year = 2024,
        month = feb,
       volume = {962},
       number = {2},
          eid = {120},
        pages = {120},
          doi = {10.3847/1538-4357/ad1591},
archivePrefix = {arXiv},
       eprint = {2312.07750},
 primaryClass = {astro-ph.GA},
       adsurl = {https://ui.adsabs.harvard.edu/abs/2024ApJ...962..120M}
}

@ARTICLE{Ziolkowska2024,
       author = {{Zi{\'o}{\l}kowska}, O. and {Smolec}, R. and {Thoul}, A. and {Farrell}, E. and {Rathour}, R. Singh and {Hocd{\'e}}, V.},
        title = "{Toward a Comprehensive Grid of Cepheid Models with MESA. I. Uncertainties of the Evolutionary Tracks of Intermediate-mass Stars}",
      journal = {\apjs},
         year = 2024,
        month = oct,
       volume = {274},
       number = {2},
          eid = {30},
        pages = {30},
          doi = {10.3847/1538-4365/ad614d},
archivePrefix = {arXiv},
       eprint = {2408.07136},
 primaryClass = {astro-ph.SR},
       adsurl = {https://ui.adsabs.harvard.edu/abs/2024ApJS..274...30Z}
}

@ARTICLE{Pilecki2024,
       author = {{Pilecki}, Bogumi{\l}},
        title = "{Fundamentalization of Periods for First- and Second-overtone Classical Cepheids}",
      journal = {\apjl},
         year = 2024,
        month = jul,
       volume = {970},
       number = {1},
          eid = {L14},
        pages = {L14},
          doi = {10.3847/2041-8213/ad5b54},
archivePrefix = {arXiv},
       eprint = {2406.18656},
 primaryClass = {astro-ph.SR},
       adsurl = {https://ui.adsabs.harvard.edu/abs/2024ApJ...970L..14P}
}

@ARTICLE{Somma2024,
       author = {{De Somma}, Giulia and {Marconi}, Marcella and {Cassisi}, Santi and {Molinaro}, Roberto},
        title = "{Stellar Pulsation and Evolution: A Combined Theoretical Renewal and Updated Models (SPECTRUM). I. Updating Radiative Opacities for Pulsation Models of Classical Cepheid and RR-Lyrae}",
      journal = {\apj},
         year = 2024,
        month = dec,
       volume = {977},
       number = {1},
          eid = {1},
        pages = {1},
          doi = {10.3847/1538-4357/ad8eb2},
       adsurl = {https://ui.adsabs.harvard.edu/abs/2024ApJ...977....1D}
}

@ARTICLE{Marconi2005,
       author = {{Marconi}, M. and {Musella}, I. and {Fiorentino}, G.},
        title = "{Cepheid Pulsation Models at Varying Metallicity and {\ensuremath{\Delta}}Y/{\ensuremath{\Delta}}Z}",
      journal = {\apj},
         year = 2005,
        month = oct,
       volume = {632},
       number = {1},
        pages = {590-610},
          doi = {10.1086/432790},
archivePrefix = {arXiv},
       eprint = {astro-ph/0506207},
 primaryClass = {astro-ph},
       adsurl = {https://ui.adsabs.harvard.edu/abs/2005ApJ...632..590M}
}

@ARTICLE{Saio1983,
       author = {{Saio}, H. and {Winget}, D.~E. and {Robinson}, E.~L.},
        title = "{Pulsation properties of DA white dwarfs - Radial mode instabilities}",
      journal = {\apj},
         year = 1983,
        month = feb,
       volume = {265},
        pages = {982-995},
          doi = {10.1086/160740},
       adsurl = {https://ui.adsabs.harvard.edu/abs/1983ApJ...265..982S}
}

@ARTICLE{Stellingwerf1982,
       author = {{Stellingwerf}, R.~F.},
        title = "{Convection in pulsating stars. I. Non linear hydro-dynamics.}",
      journal = {\apj},
         year = 1982,
        month = nov,
       volume = {262},
        pages = {330-338},
          doi = {10.1086/160425},
       adsurl = {https://ui.adsabs.harvard.edu/abs/1982ApJ...262..330S}
}

@ARTICLE{Ripepi2017,
       author = {{Ripepi}, Vincenzo and {Cioni}, Maria-Rosa L. and {Moretti}, Maria Ida and {Marconi}, Marcella and {Bekki}, Kenji and {Clementini}, Gisella and {de Grijs}, Richard and {Emerson}, Jim and {Groenewegen}, Martin A.~T. and {Ivanov}, Valentin D. and {Molinaro}, Roberto and {Muraveva}, Tatiana and {Oliveira}, Joana M. and {Piatti}, Andr{\'e}s E. and {Subramanian}, Smitha and {van Loon}, Jacco Th.},
        title = "{The VMC survey - XXV. The 3D structure of the Small Magellanic Cloud from Classical Cepheids}",
      journal = {\mnras},
         year = 2017,
        month = nov,
       volume = {472},
       number = {1},
        pages = {808-827},
          doi = {10.1093/mnras/stx2096},
archivePrefix = {arXiv},
       eprint = {1707.04500},
 primaryClass = {astro-ph.GA},
       adsurl = {https://ui.adsabs.harvard.edu/abs/2017MNRAS.472..808R}
}

@ARTICLE{Espinoza2025,
       author = {{Espinoza-Arancibia}, Felipe and {Pilecki}, Bogumi{\l}},
        title = "{A Novel q-PED Method: Precise Physical Properties of a Merger-origin Binary Cepheid OGLE-LMC-CEP-1347}",
      journal = {\apjl},
         year = 2025,
        month = mar,
       volume = {981},
       number = {2},
          eid = {L35},
        pages = {L35},
          doi = {10.3847/2041-8213/adb96b},
archivePrefix = {arXiv},
       eprint = {2501.09076},
 primaryClass = {astro-ph.SR},
       adsurl = {https://ui.adsabs.harvard.edu/abs/2025ApJ...981L..35E}
}

@ARTICLE{Salpeter1955,
       author = {{Salpeter}, Edwin E.},
        title = "{The Luminosity Function and Stellar Evolution.}",
      journal = {\apj},
         year = 1955,
        month = jan,
       volume = {121},
        pages = {161},
          doi = {10.1086/145971},
       adsurl = {https://ui.adsabs.harvard.edu/abs/1955ApJ...121..161S}
}

@ARTICLE{Costa2025,
       author = {{Costa}, G. and {Shepherd}, K.~G. and {Bressan}, A. and {Addari}, F. and {Chen}, Y. and {Fu}, X. and {Volpato}, G. and {Nguyen}, C.~T. and {Girardi}, L. and {Marigo}, P. and {Mazzi}, A. and {Pastorelli}, G. and {Trabucchi}, M. and {Bossini}, D. and {Zaggia}, S.},
        title = "{Evolutionary tracks, ejecta, and ionizing photons from intermediate-mass to very massive stars with PARSEC}",
      journal = {\aap},
         year = 2025,
        month = feb,
       volume = {694},
          eid = {A193},
        pages = {A193},
          doi = {10.1051/0004-6361/202452573},
archivePrefix = {arXiv},
       eprint = {2501.12917},
 primaryClass = {astro-ph.SR},
       adsurl = {https://ui.adsabs.harvard.edu/abs/2025A&A...694A.193C}
}

@INPROCEEDINGS{Taormina2018,
       author = {{Taormina}, M{\'o}nica and {Pilecki}, Bogumi{\l} and {Smolec}, Rados{\l}aw},
        title = "{Pulsation Theory Models for Cepheids in Eclipsing Binary Systems}",
    booktitle = {The RR Lyrae 2017 Conference. Revival of the Classical Pulsators: from Galactic Structure to Stellar Interior Diagnostics},
         year = 2018,
       editor = {{Smolec}, R. and {Kinemuchi}, K. and {Anderson}, R.~I.},
       volume = {6},
        month = jun,
        pages = {325-326},
          doi = {10.48550/arXiv.1803.10911},
archivePrefix = {arXiv},
       eprint = {1803.10911},
 primaryClass = {astro-ph.SR},
       adsurl = {https://ui.adsabs.harvard.edu/abs/2018pas6.conf..325T}
}

@ARTICLE{Deka2025,
       author = {{Deka}, M. and {Ahlborn}, F. and {Braun}, T.~A.~M. and {Weiss}, A.},
        title = "{Implications of a turbulent convection model for classical Cepheids}",
      journal = {\aap},
         year = 2025,
        month = jul,
       volume = {699},
          eid = {A351},
        pages = {A351},
          doi = {10.1051/0004-6361/202554292},
archivePrefix = {arXiv},
       eprint = {2506.04759},
 primaryClass = {astro-ph.SR},
       adsurl = {https://ui.adsabs.harvard.edu/abs/2025A&A...699A.351D}
}

@ARTICLE{Espinoza2022,
       author = {{Espinoza-Arancibia}, F. and {Catelan}, M. and {Hajdu}, G. and {Rodr{\'\i}guez-Segovia}, N. and {Boggiano}, G. and {Joachimi}, K. and {Mu{\~n}oz-L{\'o}pez}, C. and {Ordenes-Huanca}, C. and {Orquera-Rojas}, C. and {Torres}, P. and {Valenzuela-Navarro}, {\'A}.},
        title = "{Period change rates of Large Magellanic Cloud Cepheids using MESA}",
      journal = {\mnras},
         year = 2022,
        month = nov,
       volume = {517},
       number = {1},
        pages = {1538-1556},
          doi = {10.1093/mnras/stac2732},
archivePrefix = {arXiv},
       eprint = {2209.10609},
 primaryClass = {astro-ph.SR},
       adsurl = {https://ui.adsabs.harvard.edu/abs/2022MNRAS.517.1538E}
}

@ARTICLE{Pilecki2022,
       author = {{Pilecki}, Bogumi{\l} and {Thompson}, Ian B. and {Espinoza-Arancibia}, Felipe and {Anderson}, Richard I. and {Gieren}, Wolfgang and {Narloch}, Weronika and {Minniti}, Javier and {Pietrzy{\'n}ski}, Grzegorz and {Taormina}, M{\'o}nica and {Bono}, Giuseppe and {Hajdu}, Gergely},
        title = "{Discovery of a Binary-origin Classical Cepheid in a Binary System with a 59 day Orbital Period}",
      journal = {\apjl},
         year = 2022,
        month = dec,
       volume = {940},
       number = {2},
          eid = {L48},
        pages = {L48},
          doi = {10.3847/2041-8213/ac9fcc},
archivePrefix = {arXiv},
       eprint = {2212.04518},
 primaryClass = {astro-ph.SR},
       adsurl = {https://ui.adsabs.harvard.edu/abs/2022ApJ...940L..48P}
}

@ARTICLE{Pilecki2021,
       author = {{Pilecki}, Bogumi{\l} and {Pietrzy{\'n}ski}, Grzegorz and {Anderson}, Richard I. and {Gieren}, Wolfgang and {Taormina}, M{\'o}nica and {Narloch}, Weronika and {Evans}, Nancy R. and {Storm}, Jesper},
        title = "{Cepheids with Giant Companions. I. Revealing a Numerous Population of Double-lined Binary Cepheids}",
      journal = {\apj},
         year = 2021,
        month = apr,
       volume = {910},
       number = {2},
          eid = {118},
        pages = {118},
          doi = {10.3847/1538-4357/abe7e9},
archivePrefix = {arXiv},
       eprint = {2102.11302},
 primaryClass = {astro-ph.SR},
       adsurl = {https://ui.adsabs.harvard.edu/abs/2021ApJ...910..118P}
}

@ARTICLE{Madore2017,
       author = {{Madore}, Barry F. and {Freedman}, Wendy L. and {Moak}, Sandy},
        title = "{A Method for Improving Galactic Cepheid Reddenings and Distances}",
      journal = {\apj},
         year = 2017,
        month = jun,
       volume = {842},
       number = {1},
          eid = {42},
        pages = {42},
          doi = {10.3847/1538-4357/aa6e4d},
       adsurl = {https://ui.adsabs.harvard.edu/abs/2017ApJ...842...42M}
}

@ARTICLE{pilecki2024a,
       author = {{Pilecki}, Bogumi{\l} and {Thompson}, Ian B. and {Espinoza-Arancibia}, Felipe and {Hajdu}, Gergely and {Gieren}, Wolfgang and {Taormina}, M{\'o}nica and {Pietrzy{\'n}ski}, Grzegorz and {Narloch}, Weronika and {Bono}, Giuseppe and {Gallenne}, Alexandre and {Kervella}, Pierre and {Wielg{\'o}rski}, Piotr and {Zgirski}, Bart{\l}omiej and {Graczyk}, Dariusz and {Karczmarek}, Paulina and {Evans}, Nancy R.},
        title = "{Cepheids with giant companions. II. Spectroscopic confirmation of nine new double-lined binary systems composed of two Cepheids}",
      journal = {\aap},
         year = 2024,
        month = jun,
       volume = {686},
          eid = {A263},
        pages = {A263},
          doi = {10.1051/0004-6361/202349138},
archivePrefix = {arXiv},
       eprint = {2403.12390},
 primaryClass = {astro-ph.SR},
       adsurl = {https://ui.adsabs.harvard.edu/abs/2024A&A...686A.263P}
}

@ARTICLE{Kurbah2025,
       author = {{Kurbah}, Kerdaris and {Kanbur}, Shashi M. and {Deb}, Sukanta and {Bhardwaj}, Anupam and {Deka}, Mami and {Das}, Susmita and {Bhuyan}, Gautam},
        title = "{Multiwavelength study of observed and predicted pulsation properties of first overtone Cepheids in the Magellanic Clouds}",
      journal = {\mnras},
         year = 2025,
        month = aug,
       volume = {541},
       number = {3},
        pages = {2594-2618},
          doi = {10.1093/mnras/staf1023},
archivePrefix = {arXiv},
       eprint = {2506.15171},
 primaryClass = {astro-ph.GA},
       adsurl = {https://ui.adsabs.harvard.edu/abs/2025MNRAS.541.2594K}
}

@article{Pilgrim2021, doi = {10.21105/joss.03859}, url = {https://doi.org/10.21105/joss.03859}, year = {2021}, publisher = {The Open Journal}, volume = {6}, number = {68}, pages = {3859}, author = {Pilgrim, Charlie}, title = {piecewise-regression (aka segmented regression) in Python}, journal = {Journal of Open Source Software} }

@article{Muggeo2003,
author = {Muggeo, Vito M. R.},
title = {Estimating regression models with unknown break-points},
journal = {Statistics in Medicine},
volume = {22},
number = {19},
pages = {3055-3071},
doi = {https://doi.org/10.1002/sim.1545},
url = {https://onlinelibrary.wiley.com/doi/abs/10.1002/sim.1545},
eprint = {https://onlinelibrary.wiley.com/doi/pdf/10.1002/sim.1545},
year = {2003}
}

@ARTICLE{Bono2024,
       author = {{Bono}, G. and {Braga}, V.~F. and {Pietrinferni}, A.},
        title = "{Cepheids as distance indicators and stellar tracers}",
      journal = {\aapr},
         year = 2024,
        month = apr,
       volume = {32},
       number = {1},
          eid = {4},
        pages = {4},
          doi = {10.1007/s00159-024-00153-0},
archivePrefix = {arXiv},
       eprint = {2405.04893},
 primaryClass = {astro-ph.SR},
       adsurl = {https://ui.adsabs.harvard.edu/abs/2024A&ARv..32....4B}
}

@ARTICLE{Stuck2025,
       author = {{Stuck}, M. and {Pratt}, J. and {Baraffe}, I. and {Guzik}, J.~A. and {Dethero}, M. -G. and {Vlaykov}, D.~G. and {Goffrey}, T. and {Le Saux}, A.},
        title = "{Convective shells in the interior of Cepheid variable stars: Overshooting models based on hydrodynamic simulations}",
      journal = {\aap},
         year = 2025,
        month = jun,
       volume = {698},
          eid = {A304},
        pages = {A304},
          doi = {10.1051/0004-6361/202555172},
archivePrefix = {arXiv},
       eprint = {2505.04900},
 primaryClass = {astro-ph.SR},
       adsurl = {https://ui.adsabs.harvard.edu/abs/2025A&A...698A.304S}
}

@ARTICLE{Hocde2024,
       author = {{Hocd{\'e}}, V. and {Smolec}, R. and {Moskalik}, P. and {Singh Rathour}, R. and {Zi{\'o}{\l}kowska}, O.},
        title = "{Pulsation modeling of the Cepheid Y Ophiuchi with RSP/MESA. Impact of the circumstellar envelope and a high projection factor on the Baade-Wesselink method}",
      journal = {\aap},
         year = 2024,
        month = mar,
       volume = {683},
          eid = {A233},
        pages = {A233},
          doi = {10.1051/0004-6361/202348428},
archivePrefix = {arXiv},
       eprint = {2312.12046},
 primaryClass = {astro-ph.SR},
       adsurl = {https://ui.adsabs.harvard.edu/abs/2024A&A...683A.233H}
}

@ARTICLE{Marconi2024,
       author = {{Marconi}, Marcella and {De Somma}, Giulia and {Molinaro}, Roberto and {Bhardwaj}, Anupam and {Ripepi}, Vincenzo and {Musella}, Ilaria and {Sicignano}, Teresa and {Trentin}, Erasmo and {Leccia}, Silvio},
        title = "{The Hertzsprung progression of classical Cepheids in the Gaia era}",
      journal = {\mnras},
         year = 2024,
        month = apr,
       volume = {529},
       number = {4},
        pages = {4210-4233},
          doi = {10.1093/mnras/stae734},
archivePrefix = {arXiv},
       eprint = {2403.05699},
 primaryClass = {astro-ph.SR},
       adsurl = {https://ui.adsabs.harvard.edu/abs/2024MNRAS.529.4210M}
}

@INPROCEEDINGS{Buchler2009,
       author = {{Buchler}, J. Robert},
        title = "{The State of Cepheid Pulsation Theory}",
    booktitle = {Stellar Pulsation: Challenges for Theory and Observation},
         year = 2009,
       editor = {{Guzik}, Joyce Ann and {Bradley}, Paul A.},
       series = {American Institute of Physics Conference Series},
       volume = {1170},
        month = sep,
    publisher = {AIP},
        pages = {51-58},
          doi = {10.1063/1.3246556},
archivePrefix = {arXiv},
       eprint = {0907.1766},
 primaryClass = {astro-ph.SR},
       adsurl = {https://ui.adsabs.harvard.edu/abs/2009AIPC.1170...51B}
}

@ARTICLE{Breuval2025,
       author = {{Breuval}, Louise and {Anand}, Gagandeep S. and {Anderson}, Richard I. and {Beaton}, Rachael and {Bhardwaj}, Anupam and {Casertano}, Stefano and {Clementini}, Gisella and {Cruz Reyes}, Mauricio and {De Somma}, Giulia and {Groenewegen}, Martin A.~T. and {Huang}, Caroline D. and {Kervella}, Pierre and {Khan}, Saniya and {Macri}, Lucas M. and {Marconi}, Marcella and {Minniti}, Javier H. and {Riess}, Adam G. and {Ripepi}, Vincenzo and {Romaniello}, Martino and {Scolnic}, Daniel and {Trentin}, Erasmo and {Wielgorski}, Piotr and {Yuan}, Wenlong},
        title = "{Converging on the Cepheid Metallicity Dependence: Implications of Non-Standard Gaia Parallax Recalibration on Distance Measures}",
      journal = {arXiv e-prints},
         year = 2025,
        month = jul,
          eid = {arXiv:2507.15936},
        pages = {arXiv:2507.15936},
          doi = {10.48550/arXiv.2507.15936},
archivePrefix = {arXiv},
       eprint = {2507.15936},
 primaryClass = {astro-ph.GA},
       adsurl = {https://ui.adsabs.harvard.edu/abs/2025arXiv250715936B}
}

@ARTICLE{Khan2025,
       author = {{Khan}, S. and {Anderson}, R.~I. and {Ekstr{\"o}m}, S. and {Georgy}, C. and {Breuval}, L.},
        title = "{The stellar evolution perspective on the metallicity dependence of classical Cepheid Leavitt laws}",
      journal = {\aap},
         year = 2025,
        month = oct,
       volume = {702},
          eid = {A235},
        pages = {A235},
          doi = {10.1051/0004-6361/202555704},
archivePrefix = {arXiv},
       eprint = {2505.22512},
 primaryClass = {astro-ph.SR},
       adsurl = {https://ui.adsabs.harvard.edu/abs/2025A&A...702A.235K}
}

@ARTICLE{Breuval2022,
       author = {{Breuval}, Louise and {Riess}, Adam G. and {Kervella}, Pierre and {Anderson}, Richard I. and {Romaniello}, Martino},
        title = "{An Improved Calibration of the Wavelength Dependence of Metallicity on the Cepheid Leavitt Law}",
      journal = {\apj},
         year = 2022,
        month = nov,
       volume = {939},
       number = {2},
          eid = {89},
        pages = {89},
          doi = {10.3847/1538-4357/ac97e2},
archivePrefix = {arXiv},
       eprint = {2205.06280},
 primaryClass = {astro-ph.GA},
       adsurl = {https://ui.adsabs.harvard.edu/abs/2022ApJ...939...89B}
}
\begin{appendix}
\onecolumn
\section{Tables with coefficients of the red and blue edges of the IS}

\begin{table*}[h!]
\caption{Coefficients of the red and blue edges of the IS considering a break at different pulsation periods, assuming $(V-I)_{0}=\alpha (M_{I} + 3.5)+\beta$, for F and 1O together and separately. In addition, we include the coefficients for a fit without a break. $^{\star}$For F Cepheids, the break in the IS is observed at pulsation periods of $2$ days for the blue edge, and $3$ days for the red edge. $^{\star\star}$For 1O Cepheids, the break in the IS is observed at pulsation periods of $1.4$ days for the red edge, and $2$ days for the blue edge.}
\centering
\begin{tabular}{|c|c|c|c|c|c|c|c|c|}
\hline\hline
        $P$ [d]&$\alpha_{\rm blue}$&$\beta_{\rm blue}$&$\sigma_{\alpha,\rm blue}$&$\sigma_{\beta,\rm blue}$&$\alpha_{\rm red}$&$\beta_{\rm red}$&$\sigma_{\alpha,\rm red}$&$\sigma_{\beta,\rm red}$\\\hline
        \multicolumn{9}{|c|}{F and 1O Cepheids} \\\hline
        $< 2.5$ d &-0.036 & 0.456 & 0.005 & 0.006 & -0.035 & 0.709 & 0.003 & 0.005\\
        $> 2.5$ d &-0.078 & 0.493 & 0.002 & 0.001 &-0.118 & 0.744 & 0.023 & 0.013\\
        all &-0.069 & 0.489 & - & - &  -0.058 & 0.740 & - & -\\\hline
        \multicolumn{9}{|c|}{F Cepheids} \\\hline
        $< 2$ or $3$ d$^{\star}$ &0.024 & 0.434 & 0.011 & 0.015 & -0.056 & 0.749 & 0.011 & 0.012\\
        $> 2$ or $3$ d &-0.060 & 0.526 & 0.008 & 0.005 & -0.167 & 0.755 & 0.018 & 0.013\\
        all &-0.051 & 0.527 & - & - & -0.072 & 0.764 & - & - \\\hline
        \multicolumn{9}{|c|}{1O Cepheids} \\\hline
        $< 1.4$ or $2$ d$^{\star\star}$ &-0.051 & 0.449 & 0.010 & 0.011 & 0.040 & 0.530 & 0.009 & 0.016\\
        $> 1.4$ or $2$ d &-0.101 & 0.466 & 0.005 & 0.001 & -0.052 & 0.651 & 0.006 & 0.007\\
        all &-0.063 & 0.461 & - & - & -0.020 & 0.625 & - & - \\\hline\hline
\end{tabular}
\label{tab:table1}
\end{table*}

\begin{table*}[h!]
\caption{Coefficients of the red and blue edges of the IS considering a break at different pulsating periods, assuming $\log{T_{\rm eff}}=\alpha(\log {L} - 3.3)+\beta$, for F and 1O together and separately. In addition, we include the coefficients without a break. $^{\star}$ For F Cepheids, the break in the IS is observed at pulsation periods of $2$ days for the blue edge, and $3$ days for the red edge. $^{\star\star}$ For 1O Cepheids, the break in the IS is observed at pulsation periods of $1.4$ days for the red edge, and $2$ days for the blue edge.}
\centering
\begin{tabular}{|c|c|c|c|c|c|c|c|c|}
\hline\hline
        $P$ [d]&$\alpha_{\rm blue}$&$\beta_{\rm blue}$&$\sigma_{\alpha,\rm blue}$&$\sigma_{\beta,\rm blue}$&$\alpha_{\rm red}$&$\beta_{\rm red}$&$\sigma_{\alpha,\rm red}$&$\sigma_{\beta,\rm red}$\\\hline
        \multicolumn{9}{|c|}{F and 1O Cepheids} \\\hline
        $< 2.5$ d &-0.020 & 3.816 & 0.005 & 0.003 & -0.024 & 3.749 & 0.003 & 0.002\\
        $> 2.5$ d &-0.065 & 3.798 & 0.005 & 0.001 & -0.071 & 3.722 & 0.004 & 0.002\\
        all & -0.043 & 3.801 & - & - & -0.042 & 3.735 & - & - \\\hline
        \multicolumn{9}{|c|}{F Cepheids} \\\hline
        $< 2$ or $3$ d$^{\star}$ &0.014 & 3.828 & 0.006 & 0.005 & -0.040 & 3.733 & 0.009 & 0.007\\
        $> 2$ or $3$ d & -0.040 & 3.792 & 0.005 & 0.002 & -0.137 & 3.700 & 0.010 & 0.003\\
        all &-0.032 & 3.794 & - & - & -0.053 & 3.724 & - & - \\\hline
        \multicolumn{9}{|c|}{1O Cepheids} \\\hline
        $< 1.4$ or $2$ d$^{\star\star}$ & -0.031 & 3.814 & 0.006 & 0.004 & 0.023 & 3.806 & 0.007 & 0.006\\
        $> 1.4$ or $2$ d & -0.066 & 3.802 & 0.003 & 0.001 & -0.037 & 3.762 & 0.004 & 0.003\\
        all & -0.038 & 3.810 & - & - & -0.016 & 3.773 & - & - \\\hline\hline
\end{tabular}
\label{tab:table2}
\end{table*}

\end{appendix}
\end{document}